\documentclass[%
reprint,
amsmath,amssymb,
prb,
superscriptaddress
]{revtex4-1}
\usepackage{graphicx}
\usepackage{dcolumn}
\usepackage{bm}
\usepackage{hyperref}
\usepackage[mathlines]{lineno}
\usepackage[caption=false]{subfig}
\usepackage{amsmath}
\usepackage{amssymb}
\usepackage{array}
\newcommand{\beq} {\begin{equation}}
\newcommand{\eeq} {\end{equation}}
\newcommand{\bea} {\begin{eqnarray}}
\newcommand{\eea} {\end{eqnarray}}

\begin{document}

\title{Pairing in  quantum-critical systems: $T_c$, $\Delta$, and their ratio}
\author{Yi-Ming Wu}
\affiliation{School of Physics and Astronomy, University of Minnesota, Minneapolis, Minnesota 55455, USA}
\author{Artem Abanov}
\affiliation{Department of Physics, Texas A\&M University, College Station, Texas 77843, USA}
\author{Andrey V. Chubukov}
\affiliation{School of Physics and Astronomy, University of Minnesota, Minneapolis, Minnesota 55455, USA}

\date{\today}

\begin{abstract}
  We compute the ratio of the pairing gap $\Delta$ at $T=0$ and $T_c$  for a set of quantum-critical models in which the pairing interaction is mediated by a gapless boson with local  susceptibility $\chi (\Omega) \propto 1/|\Omega|^\gamma$ (the $\gamma$ model). The limit
   $\gamma = 0+$ ($\chi (\Omega) =\log {|\Omega|}$) describes color superconductivity, and models with $\gamma >0$ describe superconductivity in a metal at the onset  of charge or spin order. The  ratio $2\Delta/T_c$  has been recently computed numerically for $0<\gamma <2$ within Eliashberg theory  and was found  to increase with increasing $\gamma$
  [T-H Lee et al, arXiv:1805.10280].  We argue that the origin of the increase is the divergence of  $2\Delta/T_c$ at $\gamma =3$. We obtain an approximate analytical formula for
   $2\Delta/T_c$ for  $\gamma \leq 3$    and show that it agrees well with the numerics.  We also consider in detail the opposite limit of small $\gamma$.  Here we obtain the explicit expressions for $T_c$ and $\Delta$, including numerical prefactors.
     We show that these prefactors depend on fermionic self-energy in a rather non-trivial way.
      The ratio $2\Delta/T_c$ approaches the BCS value $3.53$ at  $\gamma \to 0$.
  \end{abstract}

\maketitle

\section{Introduction}

The
  BCS theory of superconductivity\cite{BCS1,*BCS2} is rightly considered to be one of the most elegant theoretical works of the 20th century.  Not only it explains how to obtain the energy gap in the fermionic spectrum, $\Delta$, and the transition temperature $T_c$ as functions of  material-dependent parameters,  but it also predicts that the ratio  $2 \Delta/T_c = 3.53$ is a material-independent  universal number.
 Measurements on ordinary superconductors, like aluminum, did find $2\Delta/T_c$ ratio consistent with BCS theory~\cite{Scalapino1969}
   However, in other materials, including novel superconductors,  $2\Delta/T_c$ is higher.  The two obvious reasons, particularly applicable to the cuprates, are non-s-wave superconductivity~\cite{Maiti_2011,betouras_karen}
    and pseudogap physics~\cite{Norman2005}.
     Another potential reason is the sensitivity of $2\Delta/T_c$ to strong coupling effects. They are often associated with Mott physics~\cite{Mott}, however
       a large  $2\Delta/T_c \sim 8-13$ (depending how $\Delta$ is defined, see below)  has been found in  Eliashberg calculations of phonon-mediated s-wave superconductivity~\cite{Scalapino1969,Carbotte1990,Marsiglio1991,Combescot1995,Marsiglio2008}, in the limit when,
   Debye frequency $\omega_D$ is vanishingly small, but electron-phonon interaction $g$ is finite (in this limit, both $T_c$ and $\Delta$ scale with $g$, Refs. \cite{ad,*others1,*others2,*others3}).

Phonon-mediated pairing at $\omega_D \to 0$  is a specific realization of a more generic situation when the pairing is mediated by a massless boson with
 susceptibility $\chi (q, \Omega)$, minimally coupled to fermions.  Other examples include pairing between fermions at a half-filled Landau level, when a massless boson is a gauge field with Landau overdamped propagator $\chi (q, \Omega) \propto 1/(q^2 + a |\Omega|/q)$ (e.g., Ref. \cite{Bonesteel1996})
 and pairing in a metal at the onset of an instability towards a charge or a spin  order either with $q=0$ or with a finite lattice momentum,
  ~~\cite{Abanov2001,Abanov2003,son1999,Chubukov2005,sslee,*sslee2,subir,*subir2,moon_2,max,*max2,senthil,raghu,*raghu2,*raghu3,*raghu4,*raghu5,mack,scal,*scal2,max_last,efetov,*efetov2,raghu_15,steve_sam,varma,max_2,*max_22,*max_23,strack,*strack2,Haslinger2003}
  The pairing problem in these systems is often considered within the computational scheme similar (but not identical) to the one originally used by Eliashberg in his analysis of phonon-mediated superconductivity~\cite{Eliashberg}. Namely,  the fully renormalized pairing vertex is obtained by summing up
   series of ladder diagrams, like in BCS theory but with dynamical bosonic propagator  $\chi(q, \Omega)$,  and with fermionic propagators, which include
    one-loop fermionic self-energy. The latter comes from the same fermion-boson interaction and is computed  self-consistently with the pairing vertex.   Higher-order self-energy corrections and non-ladder renormalizations of the pairing vertex are assumed to be  small  [a necessary condition is a requirement that a soft boson is a slow mode compared to a fermion, i.e., for the same  momentum, a typical bosonic frequency must be smaller than a typical fermionic frequency]. Within this approximation~\cite{xxxx},  the momentum integration in the Eliashberg  equations can be performed exactly for a given pairing symmetry~\cite{comm_new}, and the  problem reduces to the set of coupled 1D integral equations for frequency dependent pairing vertex $\Phi (\omega)$ and fermionic self-energy $\Sigma (\omega)$\cite{Scalapino1969,Carbotte1990,Abanov2001,Abanov2003,Marsiglio2008}.  For spin-singlet pairing, which we consider here,  the two equations are, in Matsubara frequencies
   \bea\label{eq:Eliashberg1}
&&\Phi(\omega_n)= \pi T \sum_{m}\frac{\Phi(\omega_m)}{\sqrt{[\omega_m+\Sigma(\omega_m)]^2+\Phi(\omega_m)^2}}\chi_L (\omega_n-\omega_m) \nonumber \\
&&\Sigma(\omega_n)= \pi T \sum_{m}\frac{\omega_m+\Sigma(\omega_m)}{\sqrt{[\omega_m+\Sigma(\omega_m)]^2+\Phi(\omega_m)^2}}
\chi_L (\omega_n-\omega_m).
\eea
Here $\chi_L (\Omega_m)$ is the effective, local, dimensionless bosonic susceptibility (it is equal to $g^2 \chi (q, \Omega_m)$ integrated over Fermi surface with form-factors for a given pairing channel $s$, $p$, $d$, etc). For electron-phonon problem at vanishing Debye frequency, $\chi_L (\Omega) = g^2/|\Omega_m|^2$.   We consider a generic model with $\chi_L (\Omega) = g^\gamma/|\Omega_m|^\gamma$ --- the $\gamma$ model.  For  a nematic and Ising-ferromagnetic critical points $\gamma =1 - D/3$, where $D$ is a spatial dimension, for antiferromagnetic critical point $\gamma = (3-D)/2$, models with other values of $\gamma$ have also been identified\cite{son1999,Chubukov2005,Altshuler1995,Wang2013,Bergeron2012}.  A similar set of equations for the frequency dependent pairing vertex and fermionic self-energy emerges in the dynamical mean-field theory (DMFT) approach, and it was argued that for DMFT analysis of a Hund metal within three-band Hubbard model for Fe-based superconductors yields $\chi_l (\Omega) \propto 1/|\Omega|^{1.2}$ in a wide  range of frequencies\cite{Stadler2015,Kotliar}.
As additional complication, the form of $\chi_L (\Omega)$ may by itself depend on $\Phi$ due to feedback from superconductivity on the bosonic propagator~\cite{Abanov1999,Eschrig2006}
This can be incorporated by treating $\gamma$ below $T_c$ as temperature-dependent parameter.

The goal of our study is to extract some new physics from the analysis of  $2\Delta/T_c$ in the $\gamma$-model.  The pairing gap $\Delta (\omega_n)$ is related to the pairing vertex as $\Delta(\omega_n)=\Phi(\omega_n)/(1+\Sigma(\omega_n)/\omega_n)$, and the  Eliashberg equation for $\Delta (\omega_n)$ is
\begin{equation}\label{eq:gap}
	\Delta(\omega_n)=\pi T \sum_{m}\frac{\Delta(\omega_m)-\frac{\omega_m}{\omega_n}\Delta(\omega_n)}{\sqrt{\omega_m^2+\Delta(\omega_m)^2}}\frac{g^\gamma}{|\omega_n-\omega_m|^\gamma}
\end{equation}
The $T_c$ is obtained as the highest temperature at which Eq. (\ref{eq:gap}) has a solution.
Note that the term with $m=n$ in the r.h.s. of (\ref{eq:gap}) (the self-action term) can be neglected due to vanishing of the numerator.   To see this more clearly, one has to add a small mass term $M$ to the interaction and take the limit $M \to 0$ only at the end of calculations.  The numerator in  (\ref{eq:gap}) vanishes at $m=n$ for  any $M$.  This vanishing is the consequence of the cancellation between the contributions to the gap equation from the renormalization of the pairing vertex and the self-energy~\cite{Millis1988,Wang2016,Abanov2008}, and it  has the same physics origin as the Anderson theorem -- the independence of $T_c$ on  non-magnetic impurities\cite{ANDERSON195926}. Indeed, the term with $m=n$ describes the scattering with zero  frequency transfer, averaged over finite momentum transfers, i.e., its role in the gap equation is equivalent to that of  elastic scattering by non-magnetic impurities.   We  remind in this regard that we consider
 spin-singlet pairing. For spin-triplet pairing, the  r.h.s. of the equation for the pairing vertex contains the extra overall factor $-1/3$, and the term with $m=n$ does not vanish, in analogy with the case when impurities are magnetic~\cite{triplet,*triplet2,*triplet3}

  For the gap $\Delta$ at $T=0$
 we will use $\Delta_0 = \Delta (\pi T)$ at the lowest temperature. One can show that   $\Delta_{0} = \Delta (\omega =0)$ on the real axis.  An alternative is to associate $\Delta$ with the frequency at which the density of states has a maximum, $\Delta_{\text{DOS}}$.   In BCS  theory $\Delta_{0} = \Delta_{\text{DOS}}$ and $2 \Delta_0/T_c =3.53$, but in the $\gamma$ model, $\Delta_{\text{DOS}} > \Delta_{0}$.  In a phonon superconductor with $\omega_D =0$,  $\Delta_{\text{DOS}} \approx (\pi/2) \Delta_0$.  This accounts for the discrepancy in reported $2\Delta/T_c$ ratio:  $2 \Delta_0/T_c \sim 8.3$, while $2\Delta_{\text{DOS}}/T_c \sim 12.9$ (Refs. \cite{Scalapino1969,Carbotte1990}).

The ratio of $2\Delta/T_c$  in the $\gamma$ model has been recently analyzed numerically for $0< \gamma \leq 2$ and was found to increase rapidly  with increasing $\gamma$\cite{Kotliar}.
 We  obtained the same result (see Fig.\ref{fig:ratio}) and also found that the increase of  $2\Delta/T_c$  accelerates at larger $\gamma$.
The goal of our work is to provide an explanation for the increase. We  argue that $2\Delta/T_c$ actually diverges at $\gamma \to 3$. The divergence  is the direct consequence of the fact that at $T=0$, when Matsubara frequencies  become  continuous variables,  the integral
 in  the r.h.s.  of the gap equation (\ref{eq:gap}) becomes singular at $\omega_m \approx \omega_n$ ($\int d x x^2/|x|^\gamma$ diverges at $\gamma \geq 3$). We obtain analytical formulas for $T_c$ and $\Delta$ near $\gamma =3$ and argue that they remain valid in a wide range of $\gamma <3$.

Another goal of our study is to analyze the opposite limit of small $\gamma$.  Here we explore the fact that  for any $\gamma  >0$, $\chi_L (\Omega) = (g/|\Omega|)^\gamma$  is a decreasing function of $\Omega$, in which case the r.h.s. of the gap equation is ultra-violet convergent, and there is no need to impose an upper cutoff in the frequency summation in (\ref{eq:gap}). We obtain the explicit expressions for $T_c$ and $\Delta$ in the small $\gamma$ limit. We show that $T_c = Q_T \omega_0$ and $\Delta = Q_\Delta \omega_0$, where $\omega_0 = g (1.4458 \gamma)^{1/\gamma}$ and $Q_T$ and $Q_\Delta$ are are numerical factors of order one.
 The scale $\omega_0$ has been identified before~\cite{max_last}
 To obtain it, one can neglect fermionic self-energy, i.e., treat fermions as free quasiparticles, like in BCS theory.
  However,  to obtain the factors $Q_T$ and $Q_\Delta$  one need to include the subleading terms in $\gamma$, and these additional terms do depend on the non-Fermi liquid  self-energy $\Sigma (\omega) \propto \omega^{1-\gamma}_m$. We show that the self-energy contributions to $Q_T$ and $Q_\Delta$ are rather non-trivial, and the result is very different from the one  in a weakly coupled Fermi liquid, where the self-energy changes the exponential factor $e^{-1/\lambda}$ into $e^{-(1 + \lambda)/\lambda} = e^{-1/\lambda}/e$ (Refs. \cite{Dolgov2005,Wang2013,Marsiglio2018}
   Still, we show that self-energy equally affects $T_c$ and $\Delta$, such that $2Q_\Delta/Q_T = 3.53$, as in BCS theory. We computed $T_c$ and $\Delta$ numerically at small $\gamma$, and found good agreement with our analytical results.

The structure of the paper is as follows. In Sec. \ref{sec:BCS}  we briefly review how $\Delta$ and $T_c$ are obtained in BCS theory. In Sec.\ref{sec:small_gamma} we study the case when $\gamma$ is small and obtain explicit formulas for  both $T_c$ and $\Delta$. The prefactors $Q_T$ and $Q_D$ are calculated both analytically and numerically. In Sec.\ref{sec:near3} we show the divergence of $\Delta$ when $\gamma\to3$.

\section{BCS theory}
\label{sec:BCS}

To set the stage for our calculations, we briefly outline how $2\Delta/T_c$ is obtained in BCS theory. Here, $\chi_L (\Omega) = \lambda$ is frequency independent, and
 $\Delta (\omega_n) = \Delta$.  The frequency sum in the gap equation diverges at large ${}\omega_m$ and one has to set the upper cutoff $\Lambda$. We then have
\bea
&&1 = \lambda \sum_{m=0}^{\frac{\Lambda}{2\pi T_c}} \frac{1}{m+1/2} = Li \left(\frac{3}{2} + \frac{\Lambda}{2\pi T_c}\right) - Li \left(\frac{1}{2}\right), \nonumber \\
&&1 = \lambda \int_0^\Lambda \frac{d \omega}{\sqrt{\omega^2 + \Delta^2}} = \lambda \log{\frac{2\Lambda}{\Delta}}, ~~~ T=0
\eea
where $Li (z) = \int_0^z dx/\log{x}$  is a logarithmic integral.
Using $Li \left(3/2 + \Lambda/(2\pi T)\right) - Li \left(1/2)\right) = \log{2 e^C \Lambda/(\pi T)}$, where  $C =0.577216$ is the Euler's constant, we immediately obtain $T_c = (2 e^C \Lambda/\pi) e^{-1/\lambda}$, $\Delta= (2 \Lambda) e^{-1/\lambda}$, and   $2\Delta/T_c = 2\pi/e^C = 3.52775$.
In Eliashberg theory with $\chi_L = \lambda$ one also has to include the self-energy $\Sigma = \lambda \omega$, Eq. (\ref{eq:Eliashberg1}),
  and then $T_c = (2 e^{C-1} \Lambda/\pi) e^{-1/\lambda}$,  $\Delta= (2 \Lambda/e) e^{-1/\lambda}$. The ratio  $2\Delta/T_c$ still remains $3.53$.

\section{Small $\gamma$}
\label{sec:small_gamma}

 We first  consider the case when $\chi_L (\Omega) = (g/|\Omega|)^\gamma$ with small but finite $\gamma$.
 As we said, for any finite $\gamma$,  the paring kernel $\chi_L (\omega_n-\omega_m)/|\omega_m|$ decreases  faster than $1/|\omega_m|$, i.e.,  frequency summation over $m$  in the r.h.s. of the gap equation (\ref{eq:gap}) converges.  This eliminates the need to introduce an upper frequency cutoff $\Lambda$, that is  $T_c$ and $\Delta$  remain finite even when $\Lambda$ is infinite.

  The small $\gamma$ limit has been considered before. Previous studies  analyzed the pairing susceptibility at $T=0$ and identified the large scale $\omega_{0} = g (1.4458\gamma)^{-1/\gamma} \gg g$, at which this  susceptibility diverges.  We  obtain $T_c \sim \omega_{0}$ explicitly by solving the linearized gap equation at a finite $T$ and non-linear gap equation at $T=0$, and find the proportionality factors.

\subsection{Calculation of $T_c$.}
\label{sec:Tc_small_gamma}

Consider first the linearized gap equation (the limit $\Delta_m \to 0$).  Neglecting the term with $m=n$ in the r.h.s. of  (\ref{eq:gap}), one can re-express (\ref{eq:gap}) as
\bea
&& \Delta_n \left(1 + \frac{{\tilde \Sigma}_n}{n+1/2}\right)  = \frac{K_T}{2} \sum_{m \neq n}\frac{\Delta_m} {|m+1/2|}\frac{1}{|n-m|^\gamma}, \nonumber \\
&& {\tilde \Sigma}_n  = \frac{K_T}{2} \sum_{m \neq n}\frac{\text{sign}(m+1/2)}{|n-m|^\gamma}
\label{eq:1}
\eea
where ${\tilde \Sigma}_n$ is the self-energy without the ``self-action'' term, and $K_T = \left(\frac{g}{2\pi T}\right)^\gamma$.
For $n=0, -1$, ${\tilde \Sigma}_0 = {\tilde \Sigma}_{-1} = 0$
for  $n \geq 1$,  ${\tilde \Sigma}_n = K_T \sum_1^n \frac{1}{m^\gamma}$, and for $n< 0$,   ${\tilde \Sigma}_{-n-1} = -  {\tilde \Sigma}_n$ (Refs.~\cite{Chubukov2012,Wang2016}).

We will see below that it will be sufficient to analyze Eq. (\ref{eq:1}) for  large Matsubara number $n$, however we will need all internal $m$. At large $n$, ${\tilde \Sigma}_n \approx K_T n^{1-\gamma}$.   Substituting this  into (\ref{eq:1}), we obtain
\beq
\Delta_n \left(1  + K_T \frac{1}{(n+1/2)^\gamma}\right)   = \frac{K_T}{2} \sum_{m \neq n}\frac{\Delta_m} {|m+1/2|}\frac{1}{|n-m|^\gamma}
\label{eq:2}
\eeq
 For internal $|m| < |n|$, the r.h.s. of (\ref{eq:2}) scales as $1/|n|^\gamma$  Substituting this dependence back into the r.h.s. of (\ref{eq:2}) we find that the summation over $m$ converges and yields  $O(1/\gamma)$.  Matching $1/|n|^\gamma$ dependence on both sides of Eq. (\ref{eq:2}), we find  $K_T \sim \gamma$, i.e., $T_c \propto g (1/\gamma)^{1/\gamma}$.

In order to find the prefactor in $T_c \propto g (1/\gamma)^{1/\gamma}$ we need to compute $K_{T}$ to the second order $\gamma $. For this we search for the solution in the form
\beq
\Delta_n = \frac{1}{|n+1/2|^\gamma} \sum_{p=0}^\infty \frac{a_p}{|n+1/2|^{\gamma p}}.
 \label{eq:4a}
 \eeq
Without loss of generality we set $a_{0}=1$, as the linear equation does not fix the overall magnitude of $\Delta_n$.
  Substituting this $\Delta_n$  into (\ref{eq:2}) and  matching the prefactors for $1/|n+1/2|^{p \gamma}$ with $p =1, 2, 3...$, we obtain  recursive relations for $a_p$:
  \beq
   a_p = - Z a_{p-1} \left(\frac{1}{p!(p+1)!} + \gamma\right) ,
  \label{eq:5}
  \eeq
  and the self-consistent condition on $Z$ (which determines $T_c (\gamma)$):
      \beq
 \frac{1}{Z} = \sum_{p=0}^\infty a_p \left(\frac{1}{p+1} +  \gamma \log{4 e^C} \right)
   \label{eq:sc}
  \eeq
 Here $Z = K_T/\gamma$. The terms $O(\gamma)$  in the r.h.s.  of Eq. (\ref{eq:5}) are due to the self-energy, which mixes  $1/|n|^{p \gamma}$ and $1/|n|^{(p+1) \gamma}$ gap components in  Eq. (\ref{eq:2}), the $O(\gamma)$ term in the r.h.s. of Eq. (\ref{eq:sc}) comes from the summation over Matsubara frequencies with $m= O(1)$.

  Solving Eq. (\ref{eq:5}) we obtain
  \beq
  a_p = (-Z)^p \left(\frac{1}{p! (p+1)!} + \gamma \frac{p+2}{3 p! (p-1)!}\right), ~~ p \geq 1
 \eeq
  and we remind that $a_0 =1$.  Substituting the expressions for $a_p$  into (\ref{eq:sc}) we find that it reduces to
 \bea
 && 1 = -\sum_{p=1}^\infty \frac{(-Z)^p}{(p!)^2} -  \gamma \log{4e^C} \sum_{p=1}^\infty \frac{(-Z)^p}{p! (p-1)!} \nonumber \\
 && - \frac{\gamma}{3} \sum_{p=2}^\infty  \frac{(-Z)^p (p+1)}{p! (p-2)!}
 \label{eq:5a}
 \eea
  The sums are expressed in terms of Bessel functions, and Eq. (\ref{eq:5a}) becomes
  \bea
  && J_0 (2 \sqrt{Z}) = \gamma \log{4e^C}  \sqrt{Z}  J_1 (2 \sqrt{Z}) - \nonumber \\
  && \frac{\gamma}{3}\left(Z^{3/2} J_1 (2 \sqrt{Z}) + Z  J_2 (2 \sqrt{Z})\right)
\label{eq:5b}
 \eea
  Without $O(\gamma)$ terms in the r.h.s, the condition on $T_c$ is $J_0 (2 \sqrt{Z}) =0$. This equation has multiple solutions, which is fundamentally important for the understanding of the phase diagram of the $\gamma$-model~\cite{Klein,Abanov_new}.
   For our current purposes, however, it is sufficient to consider only the solution with the highest $T_c$, i.e., with the smallest $Z$. The first zero of
   $J_0 (2 \sqrt{Z})$ is at  $Z=Z_0= 1.4458$.  This yields $(g/2\pi T)^\gamma = 1.4458 \gamma (1 + O(\gamma)$, i.e.,
   $T_c = Q_T \omega_0$, where, we remind, $ \omega_0 = (1.4458 \gamma)^{-1/\gamma}$  is the
     characteristic scale extracted from the analysis of the pairing susceptibility at $T=0$ (Ref. \cite{max_last}) and $Q_T$ is the prefactor $O(1)$, which we determine below.
      The large $n$ asymptotics of the corresponding eigenfunction $\Delta_n$ is
  \bea
  && \Delta_n = \frac{1}{|n|^\gamma} \left(1 - \frac{2 Z}{(2!)^2 |n|^\gamma} + \frac{3Z^2}{(3!)^2 |n|^{2\gamma}} + ...\right) \nonumber \\
  && = \frac{1}{|n|^\gamma} \sum_{m=0}^\infty \left(\frac{Z}{|n|^\gamma}\right)^m \frac{(-1)^m}{m! (m+1)!} = \nonumber \\
   && \frac{1}{(Z|n|^\gamma)^{1/2}} J_1 (2 \sqrt{Z/|n|^\gamma}).
  \label{eq:8}
  \eea

To obtain the  prefactor $Q_T$,  we need to include terms of order $\gamma$ because $(1 + \alpha \gamma)^{1/\gamma} = e^\alpha (1 + O(\gamma))$.
 For this we expand near $Z = Z_0$ using $J_0 (2 \sqrt{Z}) =  J_0 (2 \sqrt{Z_0}) - (Z-Z_0) J_1 (2 \sqrt{Z_0})/\sqrt{Z_0}$.  Substituting this expansion into
 (\ref{eq:5b}),  we obtain
 \beq
 Z = Z_0 \left[1 - \gamma \left(\log{\frac{4 e^C}{2.25978}}\right)\right]
\label{eq:6}
\eeq
Using $Z = K_T/\gamma = (g/(2\pi T))^\gamma/\gamma$ and $Z_0 = 1.4458$,  we obtain from Eq. (\ref{eq:6})
\beq
T_c = Q_T \omega_0 =  g  \left(1.4458 \gamma\right)^{-1/\gamma} \left(1 + O(\gamma)\right),
  \label{eq:7}
\eeq
  where
  \beq
 Q_T = \frac{1}{2\pi}\left(\frac{4 e^C}{2.25978}\right) = 0.5018
 \label{7_a}
  \eeq
 In  $Q_T$, the term $4 e^C$ comes from the summation over Matsubara frequencies $m = O(1)$, and the factor $2.25978$ comes from the self-energy. The $4 e^C$ term is the same as in BCS theory, but the self-energy contribution $2.25978$  is different from $e =2.71828$  in  BCS theory.
  This is because even for the smallest $\gamma$,   $T_c$ is determined by large Matsubara numbers, for which
  ${\tilde \Sigma}_n/\omega_n = K_T/|n|^\gamma$ cannot be approximated by a constant.

  \begin{figure}
  \begin{center}
    \includegraphics[width=8cm]{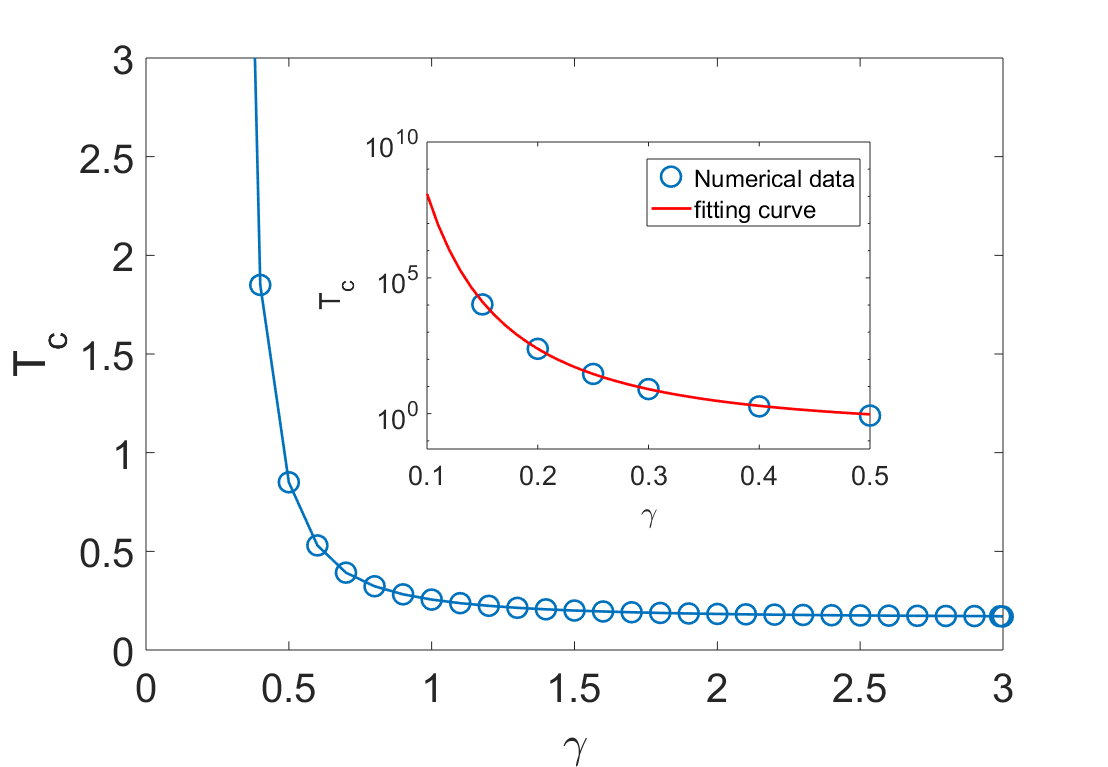}
    \caption{Numerical results of $T_c$ in units of $g$ for different $\gamma$. The inset shows our numerical data for small $\gamma$ and the red curve is the fitting curve based on \eqref{eq:7} with a prefactor $Q_T=0.4934$.}\label{fig:Tc}
  \end{center}
\end{figure}

 Our numerical results for $T_c (\gamma)$ are shown in Fig.\ref{fig:Tc}. The calculation requires care as for small $\gamma$, the solution of \eqref{eq:gap} still depends on the number of Matsubara points $M$  even when $M \sim 10^4$.  We obtained $T_c$ by  solving the gap equation on mesh of $M = m *10^3$ points  with $m$ ranging between $4$ and $16$ and then extrapolating to $M\to\infty$ (see Appendix A for details).   We see from Fig.\ref{fig:Tc} that the numerical results for $T_c/g$  are well described by
  $\left(1.4458 \gamma\right)^{-1/\gamma}$ dependence in a surprisingly broad range of $\gamma$ (roughly up to $\gamma \approx 0.5$). The numerical prefactor $Q_T$, extracted from the data, is $0.4934$, very close to $Q_T = 0.5018$ in (\ref{7_a}).  We went even further and computed the next terms in the expansion in $\gamma$. We found (see Appendix B for details) that the $O(\gamma^2)$ correction to $Z$ is quite small even for $\gamma \leq 1$, due to small prefactor.

\subsection{Calculation of the gap at $T=0$.}

We next consider the non-linear gap equation at $T=0$.  We follow the same line of reasoning as above and search for the solution for the gap at high frequencies in the form
\beq
\Delta (\omega) = \Delta f\left(\frac{\omega}{\Delta}\right).
\eeq
Substituting into the gap equation, rescaling $\omega/\Delta = {\bar \omega}$, and introducing ${\bar K} = \left(g/\Delta\right)^\gamma$, we obtain from (\ref{eq:2})
\beq
 f ({\bar \omega}) \left(1 + \frac{\bar K}{|{\bar \omega}|}\right) = \frac{\bar K}{2} \int \frac{d {\bar \omega}' f({\bar \omega}')}{\sqrt{(\bar{\omega}')^2 + f^2 (\bar{ \omega}')}} \frac{1}{|\omega- \omega'|^\gamma}
 \label{eq:8a}
 \eeq
 Like before,  we search for $f({\bar \omega})$ in the form
 \beq
 f(x) = \frac{1}{x^\gamma} \left(1 + \frac{a}{x^\gamma} + \frac{b}{x^{2\gamma}} + ... \right)
 \eeq
 For each component of $f(x)$ we represent $\int \frac{d {\bar \omega}'}{|{\bar \omega}'|^{p \gamma}} \frac{1}{\sqrt{({\bar \omega}')^2 + f^2 ({\bar \omega}')}}  \frac{1}{|\omega- \omega'|^\gamma}$
 as ${\bar A} + {\bar B}$, where
 \begin{widetext}
 \bea
 &&{\bar A} = \int \frac{d {\bar \omega}'}{|{\bar \omega}'|^{p \gamma}} \frac{1}{\sqrt{({\bar \omega}')^2 + f^2 ({\bar \omega}')}}  \frac{1}{|\omega- \omega'|^\gamma} -
 \int_1^\infty \frac{d {\bar \omega}'}{|{\bar \omega}'|^{1 + p \gamma}} \left( \frac{1}{|\omega- \omega'|^\gamma} + \frac{1}{|\omega+ \omega'|^\gamma}\right) \nonumber \\
 && {\bar B} =  \int_1^\infty \frac{d {\bar \omega}'}{|{\bar \omega}'|} \left( \frac{1}{|\omega- \omega'|^\gamma} + \frac{1}{|\omega+ \omega'|^\gamma}\right)
 \label{eq:9}
 \eea
 \end{widetext}
 In ${\bar A}$, the contribution from large ${\bar \omega}'$ cancels out, and the remaining integral reduces to ${\bar A} = 2 {\bar C}/|\omega|^\gamma$, where
 \bea
&& {\bar C} = \int \frac{d {\bar \omega}'}{|{\bar \omega}'|^{p \gamma}} \frac{1}{\sqrt{({\bar \omega}')^2 + f^2 ({\bar \omega}')}}  - \nonumber \\
&& \int_1^\infty \frac{d {\bar \omega}'}{|{\bar \omega}'|^{1 + p \gamma}}
\eea
The integral does not contain $1/\gamma$ and its leading, $\gamma$-independent piece can be computed right at $\gamma =0$, where $f ({\bar \omega}') =1$. This piece is ${\bar C} = \int_0^\infty dx/\sqrt{x^2+1} - \int_1^\infty dx/x = \log{2}$.

\begin{figure}
  \begin{center}
    \includegraphics[width=8cm]{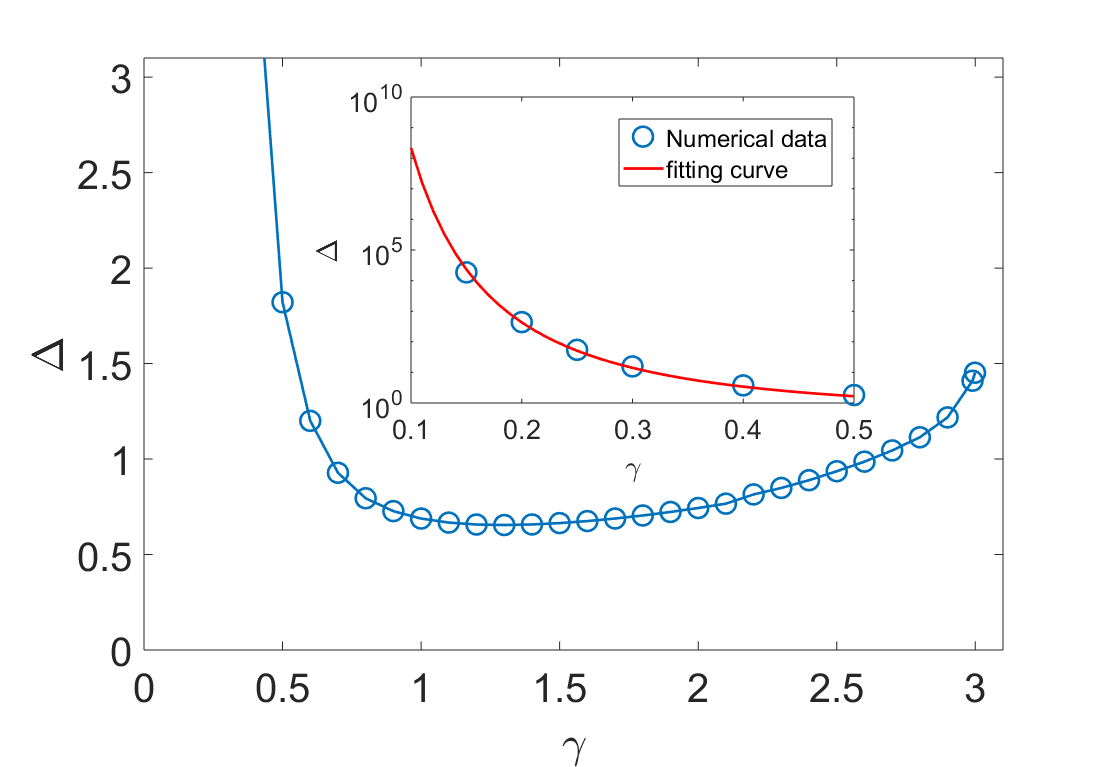}
    \caption{Numerical results for $\Delta$ (in units of $g$)  for different $\gamma$. The inset shows  numerical data for small $\gamma$,
     and the red curve is the fitting based on \eqref{eq:11}. The prefactor $Q_\Delta$, extracted from the fit, is
      $Q_\Delta=0.87$.}\label{fig:gap}
  \end{center}
\end{figure}

\begin{figure}
  \begin{center}
    \includegraphics[width=8cm]{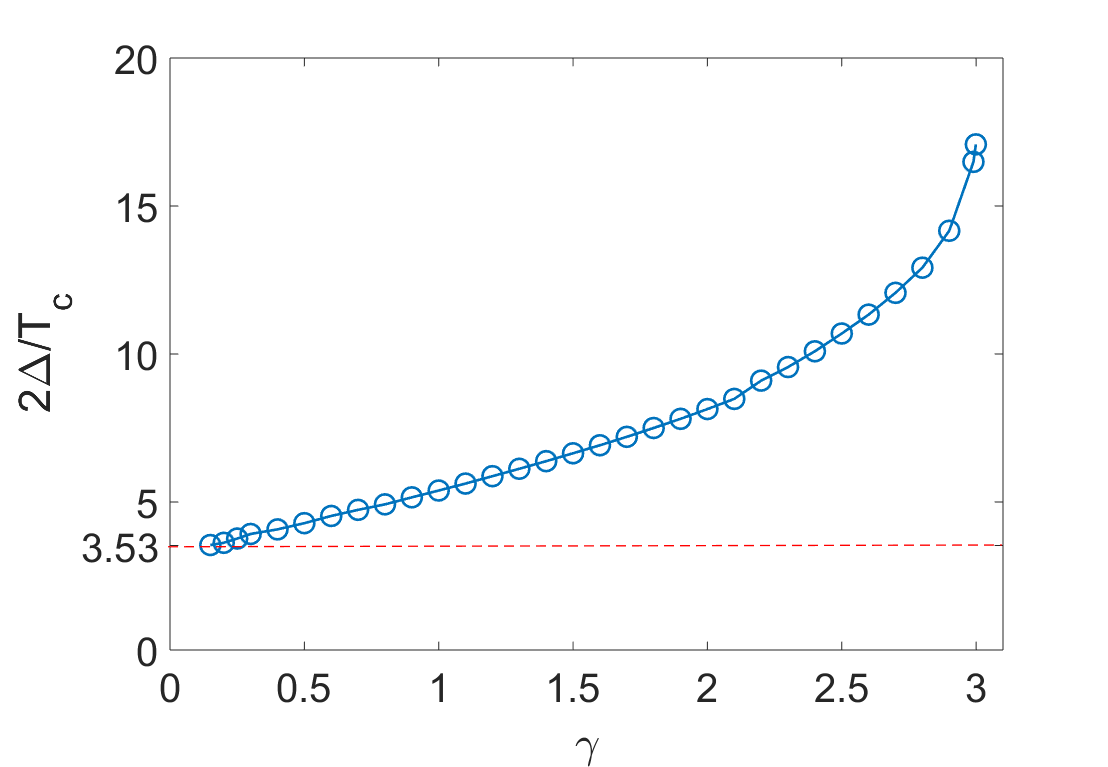}
    \caption{Numerical results for $\frac{2\Delta}{T_c}$ for different $\gamma$}\label{fig:ratio}
  \end{center}
\end{figure}

For the term ${\bar B}$,  we obtain  to order $O(\gamma )$
\bea
&& {\bar B} = \frac{2}{|\bar \omega|^\gamma} \int_1^{\bar \omega} \frac{d {\bar \omega}'}{|{\bar \omega}'|^{1 + p \gamma}} +
2 \int_{\bar \omega}^{\infty} \frac{d {\bar \omega}'}{|{\bar \omega}'|^{1 + (p+1) \gamma}}
\label{eq:10}
 \eea
 Evaluating the integrals and matching the prefactors for $1/|{\bar \omega}|^{p \gamma}$ in the r.h.s and the l.h.s of the gap equation, we obtain
 \bea
\Delta &=& Q_\Delta  \omega_0 =  Q_\Delta g  \left(1.4458 \gamma\right)^{-1/\gamma}, \nonumber \\
&&  Q_\Delta = \left(\frac{2}{2.25978}\right) = 0.885
  \label{eq:11}
\eea
Combining (\ref{eq:7}) and (\ref{eq:11}), we obtain $2 \Delta/T_c = 2 Q_D/Q_T = 2\pi/e^C = 3.53$, like in BCS theory. In the inset of Fig.\ref{fig:gap} we show our numerical results for $\Delta$ at $\gamma <0.5$.  The numerical $\Delta (\gamma)$ indeed scales with $\omega_0$.  The prefactor $Q_\Delta$, extracted from numerical data, is $0.87$, very close to the analytical $Q_\Delta = 0.885$.
 The ratio $2\Delta/T_c$ is plotted in Fig.\ref{fig:ratio}. It  clearly approaches the BCS value when $\gamma\to 0$.

\section{$T_c$ and $\Delta$ in the $\gamma$-model with  $\gamma \leq 3$.}
\label{sec:near3}

We now consider the $\gamma$ model with exponent $\gamma >1$.   We show that the ratio $2\Delta/T_c$ increases with $\gamma$ and diverges at $\gamma =3$. We argue that this divergence is the primary reason why earlier works~\cite{Scalapino1969,Carbotte1990,Combescot1995,Marsiglio2008}
 found
 a very large (but finite) $2\Delta/T_c$ by solving Eliashberg equations for a phonon superconductor with effective phonon-mediated pairing interaction
  $g^2/(\Omega^2 + \omega^2_D)$ in the limit when the Debye frequency vanishes,
  but the coupling $g$  remains  finite (the corresponds to $\gamma =2$ in our notations).

\subsection{Calculation of $T_c$}.

The onset temperature of the pairing at large $\gamma$ has been earlier analyzed by  the two of us and  collaborators~\cite{Wang2016}
  $T_c$ decreases with increasing $\gamma$ and saturates at $T_c = g/(2\pi)$ in the formal limit $\gamma \to \infty$. At large $\gamma$, the gap equation becomes local in the sense that the largest contribution to the r.h.s. of the gap equation (\ref{eq:2}) for a given $n$ comes from  $m = n \pm 1$, i.e., $\Delta_n$ is predominantly coupled to $\Delta_{n-1}$ and $\Delta_{n+1}$. This local pairing problem can be solved exactly, and
  the result is $K  \equiv  \left(g/(2\pi T_c)\right)^\gamma= 1/s$, where $s=1.1843$ is the solution of $J_{3/2 +1/s}/J_{1/2 + 1/s} = s-1$.
   Then $T_{c} = s^{1/\gamma }g/(2\pi) = \frac{g}{2\pi } \left(1 + \frac{1}{\gamma }\log{s} + ...\right)$. The dots stand for $O(1/\gamma^2)$ terms, which
     cannot be obtained within a local approach.
     The $T_c$, obtained numerically (Fig.\ref{fig:Tc}),  indeed saturates at $g/(2\pi)$ at large $\gamma$ and is actually rather close to this value for all $\gamma >1$.

  \subsection{Calculation of $\Delta$ at $T=0$}

  It is convenient to write the gap equation at $T=0$ as
  \bea
 && \Delta (\omega) \omega = \frac{g^\gamma}{2} \int \frac{d \omega'}{\sqrt{(\omega')^2 + \Delta^2 (\omega')}} \frac{\Delta (\omega') \omega - \Delta (\omega) \omega'}{|\omega - \omega'|^\gamma} = \nonumber \\
 &&  \frac{g^\gamma}{2} \int d \omega' \frac{\Delta (\omega')}{\sqrt{(\omega')^2 + \Delta^2 (\omega')}} \frac{\text{sign} (\omega - \omega')}{|\omega - \omega'|^\gamma} + \nonumber \\
 &&  \frac{g^\gamma}{2}  \int d \omega' \frac{\omega'}{\sqrt{(\omega')^2 + \Delta^2 (\omega')}} \frac{\Delta (\omega') -  \Delta (\omega)}{|\omega - \omega'|^\gamma}
\label{eq:12}
\eea
The first contribution to the r.h.s. of (\ref{eq:12}) can be re-expressed by shifting the integration variable as
\beq
\frac{g^\gamma}{2}\int_{0}^{\infty}\frac{dx}{x^{\gamma-1}}\left(\frac{1}{\sqrt{1+\left(\frac{x-\omega}{\Delta (x-\omega)}\right)^2}}-\frac{1}{\sqrt{1+\left(\frac{x+\omega}{\Delta (x+\omega)}\right)^2}}\right)
\label{eq:14}
\eeq
The second contribution can be re-expressed by collecting the terms with positive and negative $\omega'$ as
\bea
&&\frac{g^\gamma}{2}\int_{0}^{\infty}\frac{d\omega' \omega'}{\sqrt{(\omega')^2 + \Delta^2 (\omega')}} \times \nonumber \\
&& \left(\Delta (\omega') - \Delta (\omega)\right) \left(\frac{1}{|\omega - \omega'|^\gamma} +\frac{1}{|\omega + \omega'|^\gamma}\right)
\label{eq:15}
\eea
Both contributions have infra-red  divergencies $\int dx/x^{\gamma -2}$  at $\gamma >3$, as one can easily verify. However, the integral in (\ref{eq:14}),  diverges already if we approximate the gap $\Delta (\omega)$ as a constant $\Delta$ at low frequencies, while in the second contribution, Eq. (\ref{eq:15}), the divergent piece  contains $\partial^2{\Delta (\omega)}/\partial \omega^2$. We assume that this second contribution is smaller  and focus on the first one. We  set external $\omega$ in (\ref{eq:12}) and approximate its l.h.s. as
 $\Delta \omega$, where $\Delta = \Delta (\omega =0)$. The equation on $\Delta$ is then obtained by expanding in Eq. (\ref{eq:14})  to linear order  in $\omega$. Neglecting again the derivatives of $\Delta (\omega)$ we obtain
\beq
\Delta = \int_0^\infty \frac{dx}{x^{\gamma-2}} \frac{\Delta (x)}{(\Delta^2_x + x^2)^{3/2}}
\label{eq:16}
\eeq
For $\gamma$ close to $3$ the integral is determined by small $x$, and we can approximate $\Delta (x)$ by $\Delta$.  The remaining integration over $x$ can be carried out exactly, and we obtain
\beq
\Delta = g \left(\frac{\Gamma\left(\frac{\gamma}{2}\right) \Gamma\left(\frac{3-\gamma}{2}\right)}{\sqrt{\pi}}\right)^{\frac{1}{\gamma}}
\label{eq:17}
\eeq
When  $\gamma$ approaches $3$, $\Delta$ diverges as $\left(1/(3-\gamma)\right)^{1/3}$. For $\gamma =2$, Eq. (\ref{eq:17}) yields $\Delta =g$.
We note in passing that $\Delta$ given by (\ref{eq:17}) also diverges as $(1/\gamma)^{1/\gamma}$ at small $\gamma$, however the assumption that the integral in (\ref{eq:14}) is determined by small $x$ obviously does not work there.

\begin{figure}
  \begin{center}
    \includegraphics[width=4.2cm]{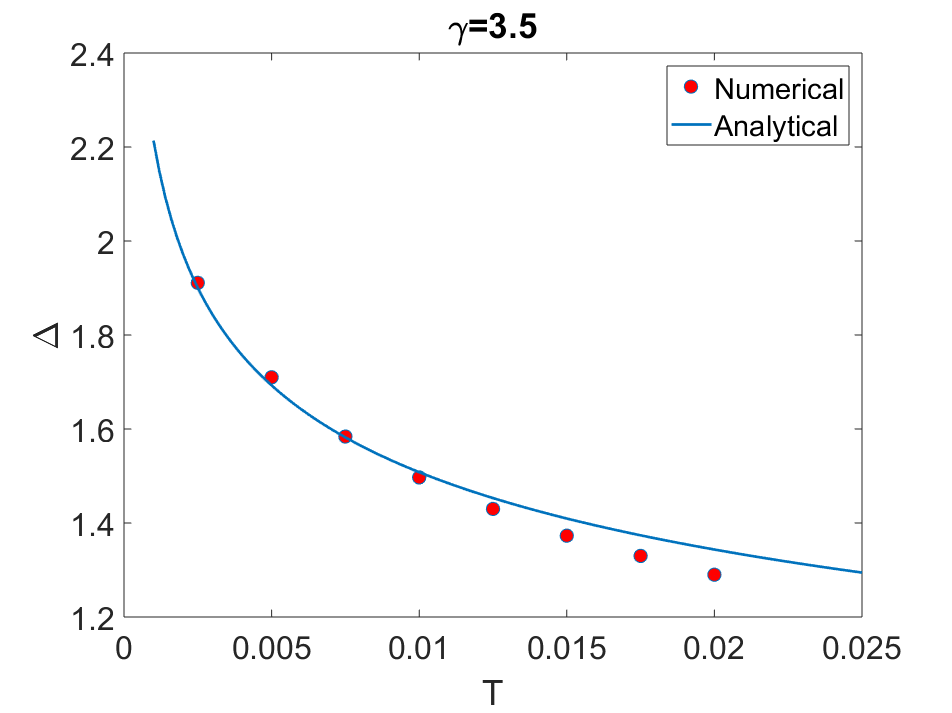}
    \includegraphics[width=4.2cm]{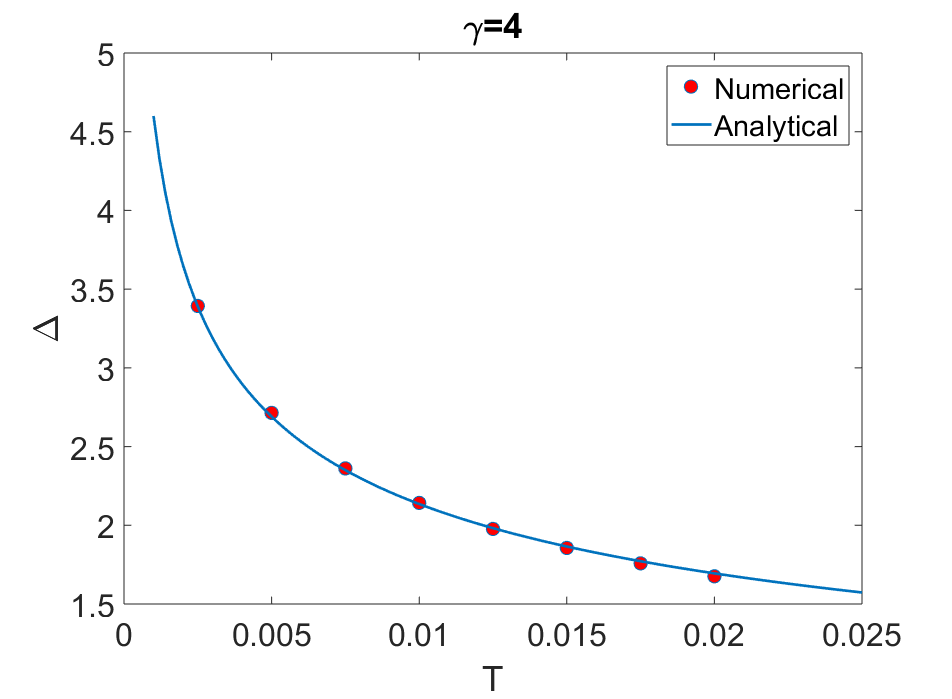}
    \caption{The behavior of $\Delta (\pi T)$ vs $T$  for $\gamma =3.5$ and $\gamma =4$.  The red dots are numerical results and the blue lines are the scaling forms $\Delta\sim \left(\frac{1}{T}\right)^{\frac{\gamma-3}{3}}$.}\label{fig:Deltadivergence}
  \end{center}
\end{figure}

As independent verification, we computed $\Delta (\pi T)$ at a finite temperature for $\gamma >3$ and indeed found that it diverges as $T \to 0$.
We show the results in Fig.\ref{fig:Deltadivergence} along with $\Delta (\pi T) \propto (1/T)^{\frac{\gamma-3}{3}}$ obtained by a straightforward
scaling analysis.   We see from Fig.\ref{fig:Deltadivergence} that numerical results reproduce this behavior quite well.

Combining $T_c \approx g/(2\pi)$ and Eq. (\ref{eq:17}) we obtain, for $\gamma \leq 3$,
\beq
\frac{2\Delta}{T_c} = 4\pi \left(\frac{\Gamma\left(\frac{\gamma}{2}\right) \Gamma\left(\frac{3-\gamma}{2}\right)}{\sqrt{\pi}}\right)^{\frac{1}{\gamma}}
\label{eq:18}
\eeq
Near $\gamma =3$,
\beq
\frac{2\Delta}{T_c} \approx 4 \pi \left(\frac{1}{3-\gamma}\right)^{1/3}.
 \label{eq:19}
 \eeq

 In  Figs.  \ref{fig:Tc}, \ref{fig:gap}, and \ref{fig:ratio} we show the numerical results for  $2\Delta/T_c$ in the full range of $\gamma$. We see that $T_c$ monotonically decreases with increasing $\gamma$ and saturates at $T_c = g/(2\pi)$ at large $\gamma$, while $\Delta$ is non-monotonic -- it diverges at $\gamma \to 0$ and $\gamma \to 3$ and has a minimum at $\gamma \approx 1$. The ratio $2\Delta/T_c$ monotonically increases with increasing $\gamma$ and diverges at $\gamma =3$.  At $\gamma =2$, $2\Delta/T_c = 8.3$, is already quite large, consistent with earlier works~~\cite{Scalapino1969,Carbotte1990,Combescot1995,Marsiglio2008}
  We see that the  large $2\Delta/T_c$ for $\gamma =2$ reflects the fact that at this $\gamma$  $2\Delta/T_c$ already accelerates towards the divergence at $\gamma =3$.

\section{Conclusions}

In this paper we analyzed superconducting $T_c$ and $2\Delta/T_c$ ratio in a metal at the verge of an instability towards a spin or a charge order. Near the instability, the dominant interaction between fermions is the exchange of soft bosonic fluctuations of spin or charge order parameter. In spatial dimension $D \leq 3$ this interaction gives rise to a non-Fermi liquid behavior either on a whole Fermi surface or in hot regions, but also provides a strong attraction in at least one pairing channel.
 We considered a subset of such systems, in which soft bosons can be regarded as slow modes compared to electrons, and the pairing can be treated
 within Eliashberg theory with an effective local interaction $\chi_L (\Omega_m) = \left(g/|\Omega_m|\right)^\gamma$ (the $\gamma$ model).
 The same effective theory emerges for the pairing between fermions at the half-filled Landau level and in models studied within DMFT.

 The $\gamma$ model with $\gamma =2$ describes electron-phonon superconductivity in the special limit when Debye frequency vanishes but
 fermion-boson coupling $g$  remains finite, i.e., the boson-mediated interaction is $(g/|\Omega|)^2$.  This problem has been extensively studied in the past~~\cite{Scalapino1969,ad,*others1,*others2,*others3,Carbotte1990,Combescot1995,Marsiglio2008}.
 It was well established that $T_c \approx 0.18 g$ and $\Delta (0) \approx 0.75 g$ remain finite, but their ratio $2\Delta/T_c \approx 8.3$ is much larger than in BCS theory.  T-H Lee et al recently analyzed numerically  $2\Delta/T_c$ in the $\gamma$-model for $\gamma <2$
  (Ref. \cite{Kotliar})
   and found that the ratio monotonically increases with increasing $\gamma$.

 One goal of our work was to provide an explanation for this increase. We  considered a larger range of  $\gamma\leq 3$ and found that
   $2\Delta (0) /T_c$ diverges at $\gamma \to 3$ ($T_c$ remains finite in this limit, but $\Delta_0$ diverges).
     We obtained analytical formulas for $T_c$ and $\Delta$ near $\gamma =3$ and argued that they remain valid in a relatively wide range of $\gamma <3$. We also computed $T_c$ and $\Delta$ numerically and found good agreement between  analytical and numerical results.

Another goal of our work  was to analyze the opposite limit of small $\gamma$.  Here we obtained the exact expressions for $T_c$ and $\Delta$ with
 numerical prefactors.  We emphasize that for any non-zero $\gamma$, the normal state self-energy has a  non-Fermi liquid form at small frequencies, and non-Fermi-liquid behavior does affect the values of $T_c$ and $\Delta$.

  \begin{figure*}
  \begin{center}
    \includegraphics[width=5cm]{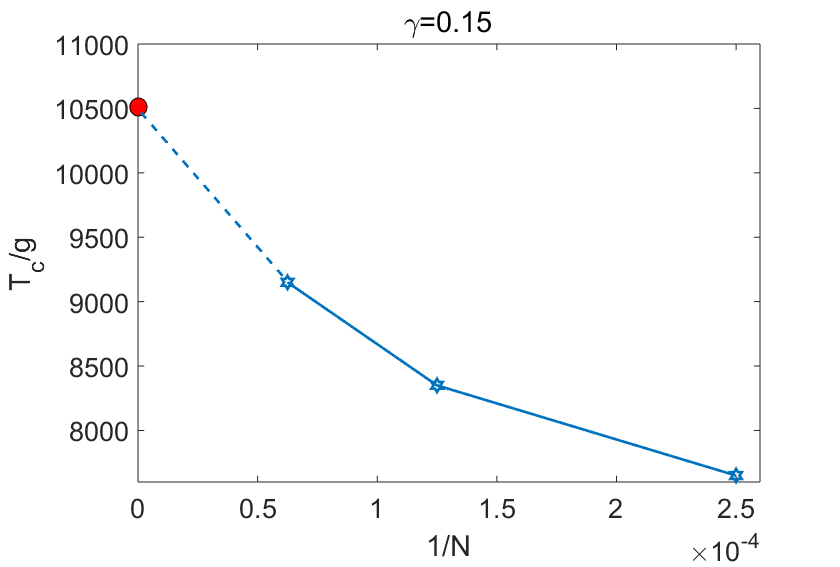}\includegraphics[width=5cm]{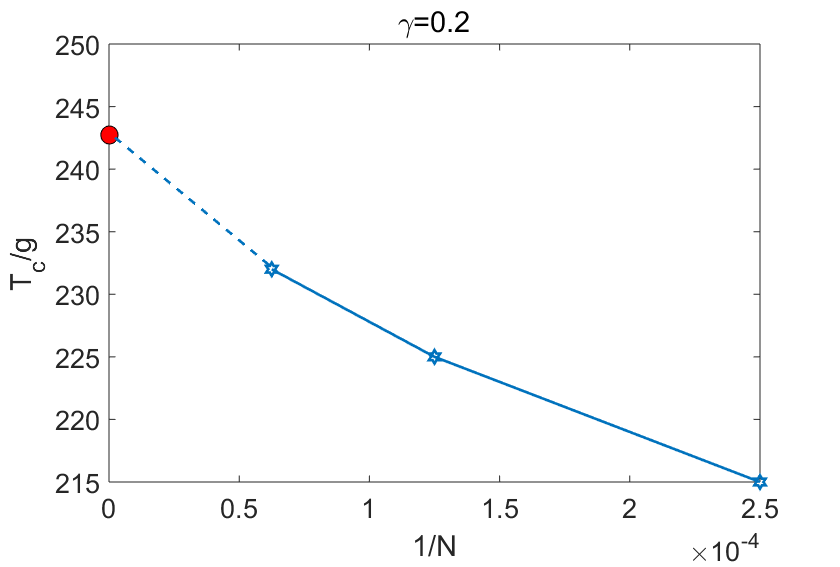}\includegraphics[width=5cm]{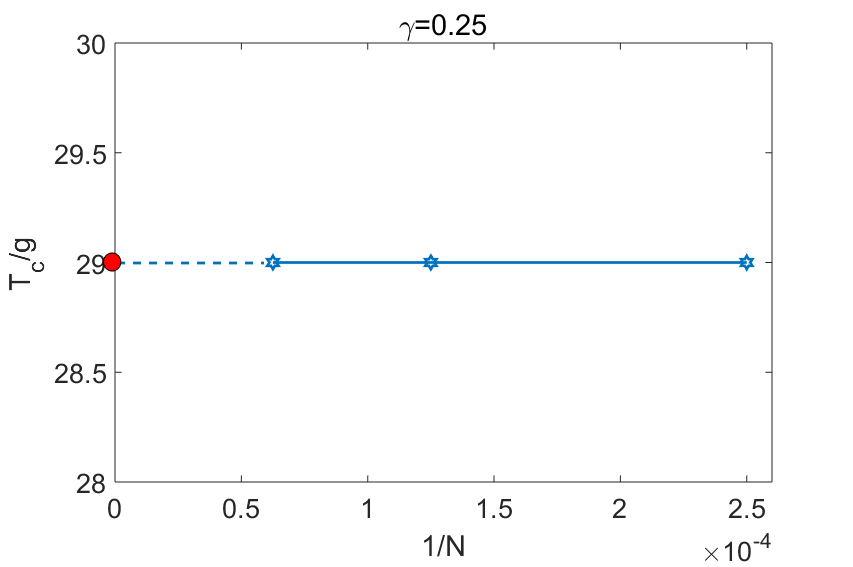}\\
    \includegraphics[width=5cm]{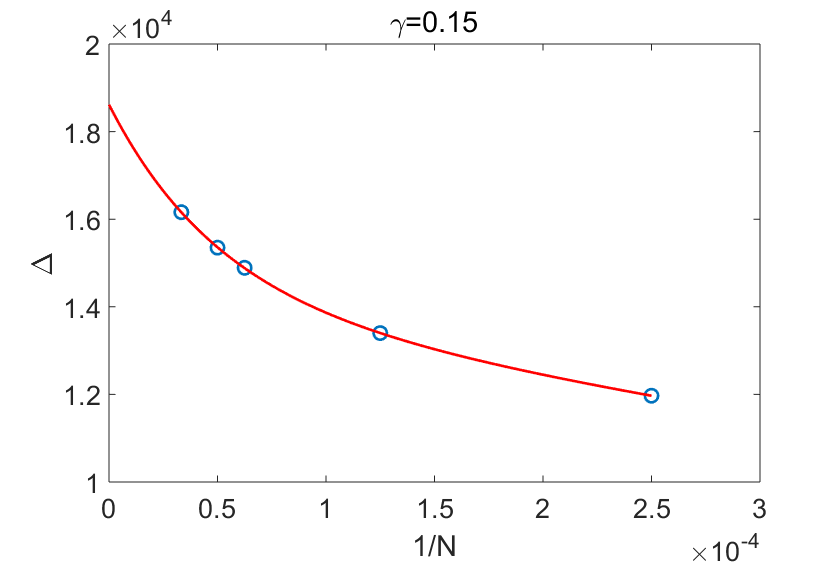}\includegraphics[width=5cm]{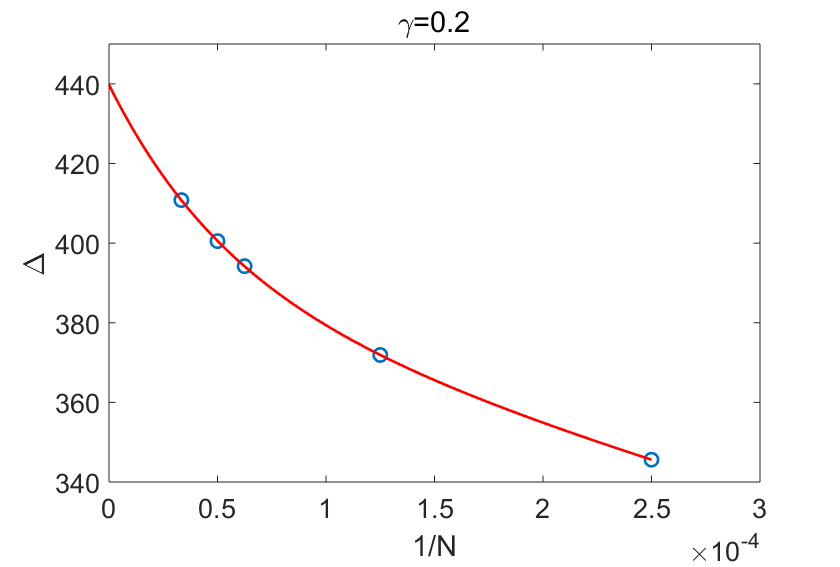}\includegraphics[width=5cm]{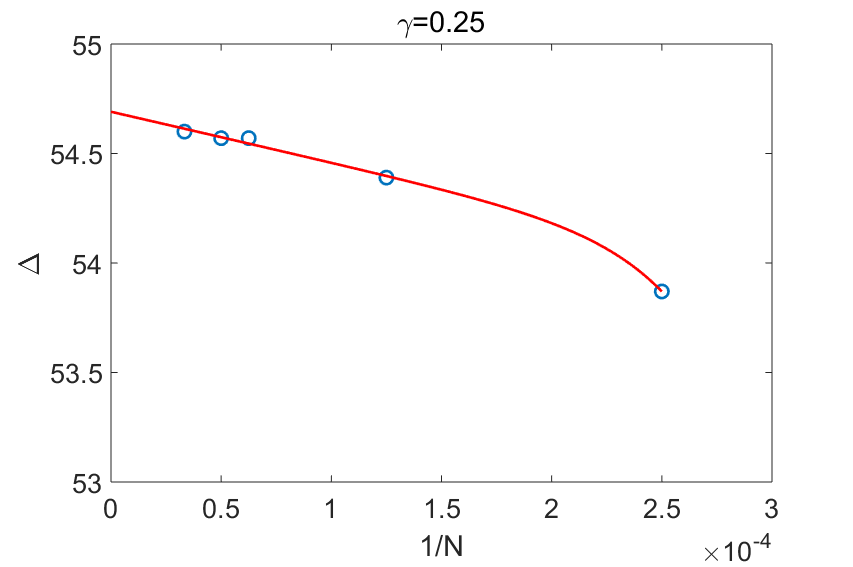}
  \end{center}
   \caption{Extrapolation procedure for $T_c$  and $\Delta$ (in units of $g$) for $\gamma=0.15$, $0.2$ and $0.25$. By extrapolating to $M\to\infty$, we obtained $Tc=10500$, $243$, and $29$, respectively, and $\Delta=18620$, $440$, and $54.6$ respectively. The red lines for $\Delta (1/N)$  are double exponential fits ($a_1 e^{-\alpha_1/N} + a_2 e^{-\alpha_2/N}$).}
  \label{fig:appendix}
\end{figure*}

   A word of caution.  In our analysis we focused on the solution of the linearized gap equation with the highest $T_c$ and on the "conventional" solution of the non-linear equation at $T=0$, for which  $\Delta (\omega_m)$ is a regular function of frequency  with no sign changes.  There exist other solutions of the gap equation, for which $\Delta (\omega_m)$ oscillates. For $\gamma <2$, there is little doubt that the conventional solution with no-nodal $\Delta (\omega_m)$ has the  largest condensation energy.
      However, for $\gamma >2$, it is possible that the largest condensation energy is for an unconventional solution with oscillating $\Delta (\omega_m)$ (Ref.\cite{Abanov_new}).
      This would affect the ratio of $2\Delta/T_c$.    Still, even if this is the case, our analysis is applicable to $\gamma \leq 2$, and it explains
       why $2\Delta/T_c$ increases with $\gamma$.
   Also, $T_c$ in our analysis is the onset temperature for the pairing instability.  In the absence of fluctuations it coincides with the actual  superconducting $T_c$, but when fluctuations are present, the actual $T_{c,act}$  likely gets smaller, while our mean-field $T_c$ marks the onset of pseudogap behavior. Our $2\Delta/T_c$ should then be understood as the ratio of the gap at $T=0$ to the onset temperature for pseudogap behavior.
   And, finally, in our analysis we neglected the feedback from the gap opening on the form of $\chi_L (\Omega)$ (e.g., the development of the resonant peak in the spin-fluctuation propagator due to the opening of $d-$wave or $s^{+-}$  gap).
  Within the $\gamma$ model this last effect can be captured by allowing $\gamma$ to vary with $T$ below $T_c$ towards a smaller value.

\begin{acknowledgments}
We thank S-L Drechsler,G. Kotliar, T-H Lee, F. Marsiglio, H. Miao, and  Y. Wang for fruitful discussions. The work by YW and AVC was supported by NSF-DMR-1523036.
\end{acknowledgments}

\section{Appendix}

\subsection{ Details of numerical calculations at  small $\gamma$}
The  results of numerical calculations of $T_c$ and $\Delta$ at small $\gamma$ are presented in the insets in Fig.\ref{fig:Tc} and Fig.\ref{fig:gap}. The  analysis requires care as at small $\gamma$ the numerical results depend on the number of Matsubara points $M$, and to obtain  reliable results  one should properly extrapolate to $M =\infty$   We solved the linearized gap equation on the sets of   $M= 4000$, $8000$ and $16000$, identified $T_c$ with the temperature when the largest eigenvalue crosses $1$, and extrapolated the results to $M = \infty$. We show the extrapolation procedure in Fig \ref{fig:appendix}a. In our calculation of $\Delta (\pi T)$ as the solution of the non-linear gap equation,  we used the fact that $\Delta$ rather quickly saturates below $T_c$, set
 the temperature to be $0.3T_c$, computed $\Delta (\pi T)$ for  $M=4000$, $8000$, $16000$, $20000$, and $30000$ Matubara points, and then extrapolated to $M = \infty$.  We show the extrapolation procedure in Fig. \ref{fig:appendix}b.

\subsection{ The calculation of $T_c$ at small $\gamma$ to order $O(\gamma^2)$}.

In this subsection we extend our analysis from Sec. \ref{sec:Tc_small_gamma} to include terms of next order in $\gamma$. The specific goal here is to understand whether there corrections are small for $\gamma =0.2$, which is the smallest $\gamma$ for which the comparison between analytic and numerical data is possible.

 The calculations follow the same path as the ones we reported in  Sec. \ref{sec:Tc_small_gamma} in the main text, i.e., we write $\Delta$ as the sum of partial components  $\Delta_n = (1/|n+1/2|^\gamma) \sum_{p=0}^\infty a_p/|n+1/2|^{\gamma p}$,  like in Eq. (\ref{eq:5}), obtain recursive relations for $a_p$, and obtain $T_c$ from self-consistent equation on $Z = K_T/\gamma =  (g/(2\pi T)^\gamma/\gamma$. However, at each step we extend the analysis to next order in $\gamma$. We skip the details of the calculations are report the results.
 The recursive relation are
  \beq
   a_p = - Z a_{p-1} \left(\frac{1}{p!(p+1)!} + \gamma^*\right)
  \label{b_3}
  \eeq
  where $\gamma^* = \gamma(1 + 0.165 \gamma)$.
   The self-consistent condition on $Z$ is
\bea
&& \frac{1}{Z} = \nonumber \\
&& \sum_{p=0}^\infty a_p \left(\frac{1}{p+1} +  \gamma \log{4 e^C} + 1.353 \gamma^2 (p+1)   \right)
   \label{b_4}
  \eea
 The solution of the recursive relation (\ref{b_3}) to order $\gamma^2$ is
 \bea
 &&a_p =  (-Z)^p  \times \label{b_5} \\
 &&  \left(\frac{1}{p! (p+1)!} + \gamma^* \frac{p+2}{3p! (p-1)!} + \gamma^2 \frac{b_p}{90 (p+1)! (p-2)!} \right) \nonumber
   \eea
  where
  \beq
  b_p = 6 +31 p + 16 p^2 +26 p^3 + 5p^4
  \label{b_6}
  \eeq
  The last term in the self-consistency equation (\ref{b_4})   is already of order $\gamma^2$,  and it can be computed
   using the leading, $\gamma$-independent terms in $a_p$.  The corresponding sum over $p$ reduces to
   \beq
   \sum_{p=0}^\infty \frac{(-Z)^p}{(p!)^2} = J_0 (2 \sqrt{Z})
   \eeq
    At $T=T_c$, $ J_0 (2 \sqrt{Z})$ is by itself of order $\gamma$, hence this term is actually of order $\gamma^3$ and can be neglected. Evaluating the remaining sums analytically and numerically, we obtained
    \bea
  && J_0 (2 \sqrt{Z}) = \gamma \log{4e^C}  \sqrt{Z}  J_1 (2 \sqrt{Z}) - \nonumber \\
  && \frac{\gamma}{3}\left(Z^{3/2} J_1 (2 \sqrt{Z}) + Z  J_2 (2 \sqrt{Z})\right) -0.30246 \gamma^2
\label{b_5a}
 \eea
    The  solution of (\ref{b_5a}) to order $\gamma^2$ is
    \beq
    Z = Z_0 \left(1 - \gamma \log{3.15265} + 0.036 \gamma^2\right),
    \label{b_7}
    \eeq
    where, we remind $Z_0 = 1.4458$ is the smallest solution of  $J_0 (2 \sqrt{Z})=0$.  We see that the $\gamma^2$ term has a very small prefactor.
    Hence the critical value of $K_T$ is only weakly affected by the $O(\gamma)^2$ term.

\bibliography{ref}

\begin{thebibliography}{68}%
\makeatletter
\providecommand \@ifxundefined [1]{%
 \@ifx{#1\undefined}
}%
\providecommand \@ifnum [1]{%
 \ifnum #1\expandafter \@firstoftwo
 \else \expandafter \@secondoftwo
 \fi
}%
\providecommand \@ifx [1]{%
 \ifx #1\expandafter \@firstoftwo
 \else \expandafter \@secondoftwo
 \fi
}%
\providecommand \natexlab [1]{#1}%
\providecommand \enquote  [1]{``#1''}%
\providecommand \bibnamefont  [1]{#1}%
\providecommand \bibfnamefont [1]{#1}%
\providecommand \citenamefont [1]{#1}%
\providecommand \href@noop [0]{\@secondoftwo}%
\providecommand \href [0]{\begingroup \@sanitize@url \@href}%
\providecommand \@href[1]{\@@startlink{#1}\@@href}%
\providecommand \@@href[1]{\endgroup#1\@@endlink}%
\providecommand \@sanitize@url [0]{\catcode `\\12\catcode `\$12\catcode
  `\&12\catcode `\#12\catcode `\^12\catcode `\_12\catcode `\%12\relax}%
\providecommand \@@startlink[1]{}%
\providecommand \@@endlink[0]{}%
\providecommand \url  [0]{\begingroup\@sanitize@url \@url }%
\providecommand \@url [1]{\endgroup\@href {#1}{\urlprefix }}%
\providecommand \urlprefix  [0]{URL }%
\providecommand \Eprint [0]{\href }%
\providecommand \doibase [0]{http://dx.doi.org/}%
\providecommand \selectlanguage [0]{\@gobble}%
\providecommand \bibinfo  [0]{\@secondoftwo}%
\providecommand \bibfield  [0]{\@secondoftwo}%
\providecommand \translation [1]{[#1]}%
\providecommand \BibitemOpen [0]{}%
\providecommand \bibitemStop [0]{}%
\providecommand \bibitemNoStop [0]{.\EOS\space}%
\providecommand \EOS [0]{\spacefactor3000\relax}%
\providecommand \BibitemShut  [1]{\csname bibitem#1\endcsname}%
\let\auto@bib@innerbib\@empty
\bibitem [{\citenamefont {Bardeen}\ \emph
  {et~al.}(1957{\natexlab{a}})\citenamefont {Bardeen}, \citenamefont {Cooper},\
  and\ \citenamefont {Schrieffer}}]{BCS1}%
  \BibitemOpen
  \bibfield  {author} {\bibinfo {author} {\bibfnamefont {J.}~\bibnamefont
  {Bardeen}}, \bibinfo {author} {\bibfnamefont {L.~N.}\ \bibnamefont {Cooper}},
  \ and\ \bibinfo {author} {\bibfnamefont {J.~R.}\ \bibnamefont {Schrieffer}},\
  }\href {\doibase 10.1103/PhysRev.106.162} {\bibfield  {journal} {\bibinfo
  {journal} {Phys. Rev.}\ }\textbf {\bibinfo {volume} {106}},\ \bibinfo {pages}
  {162} (\bibinfo {year} {1957}{\natexlab{a}})}\BibitemShut {NoStop}%
\bibitem [{\citenamefont {Bardeen}\ \emph
  {et~al.}(1957{\natexlab{b}})\citenamefont {Bardeen}, \citenamefont {Cooper},\
  and\ \citenamefont {Schrieffer}}]{BCS2}%
  \BibitemOpen
  \bibfield  {author} {\bibinfo {author} {\bibfnamefont {J.}~\bibnamefont
  {Bardeen}}, \bibinfo {author} {\bibfnamefont {L.~N.}\ \bibnamefont {Cooper}},
  \ and\ \bibinfo {author} {\bibfnamefont {J.~R.}\ \bibnamefont {Schrieffer}},\
  }\href {\doibase 10.1103/PhysRev.108.1175} {\bibfield  {journal} {\bibinfo
  {journal} {Phys. Rev.}\ }\textbf {\bibinfo {volume} {108}},\ \bibinfo {pages}
  {1175} (\bibinfo {year} {1957}{\natexlab{b}})}\BibitemShut {NoStop}%
\bibitem [{\citenamefont {Scalapino}(1969)}]{Scalapino1969}%
  \BibitemOpen
  \bibfield  {author} {\bibinfo {author} {\bibfnamefont {D.}~\bibnamefont
  {Scalapino}},\ }in\ \href@noop {} {\emph {\bibinfo {booktitle}
  {Superconductivity}}},\ \bibinfo {editor} {edited by\ \bibinfo {editor}
  {\bibfnamefont {R.}~\bibnamefont {Parks}}}\ (\bibinfo  {publisher} {CRC
  Press},\ \bibinfo {year} {1969})\BibitemShut {NoStop}%
\bibitem [{\citenamefont {Maiti}\ and\ \citenamefont
  {Chubukov}(2011)}]{Maiti_2011}%
  \BibitemOpen
  \bibfield  {author} {\bibinfo {author} {\bibfnamefont {S.}~\bibnamefont
  {Maiti}}\ and\ \bibinfo {author} {\bibfnamefont {A.~V.}\ \bibnamefont
  {Chubukov}},\ }\href {\doibase 10.1103/PhysRevB.83.220508} {\bibfield
  {journal} {\bibinfo  {journal} {Phys. Rev. B}\ }\textbf {\bibinfo {volume}
  {83}},\ \bibinfo {pages} {220508} (\bibinfo {year} {2011})}\BibitemShut
  {NoStop}%
\bibitem [{\citenamefont {Musaelian}\ \emph {et~al.}(1996)\citenamefont
  {Musaelian}, \citenamefont {Betouras}, \citenamefont {Chubukov},\ and\
  \citenamefont {Joynt}}]{betouras_karen}%
  \BibitemOpen
  \bibfield  {author} {\bibinfo {author} {\bibfnamefont {K.~A.}\ \bibnamefont
  {Musaelian}}, \bibinfo {author} {\bibfnamefont {J.}~\bibnamefont {Betouras}},
  \bibinfo {author} {\bibfnamefont {A.~V.}\ \bibnamefont {Chubukov}}, \ and\
  \bibinfo {author} {\bibfnamefont {R.}~\bibnamefont {Joynt}},\ }\href
  {\doibase 10.1103/PhysRevB.53.3598} {\bibfield  {journal} {\bibinfo
  {journal} {Phys. Rev. B}\ }\textbf {\bibinfo {volume} {53}},\ \bibinfo
  {pages} {3598} (\bibinfo {year} {1996})}\BibitemShut {NoStop}%
\bibitem [{\citenamefont {Norman}\ \emph {et~al.}(2005)\citenamefont {Norman},
  \citenamefont {Pines},\ and\ \citenamefont {Kallin}}]{Norman2005}%
  \BibitemOpen
  \bibfield  {author} {\bibinfo {author} {\bibfnamefont {M.~R.}\ \bibnamefont
  {Norman}}, \bibinfo {author} {\bibfnamefont {D.}~\bibnamefont {Pines}}, \
  and\ \bibinfo {author} {\bibfnamefont {C.}~\bibnamefont {Kallin}},\ }\href
  {\doibase 10.1080/00018730500459906} {\bibfield  {journal} {\bibinfo
  {journal} {Advances in Physics}\ }\textbf {\bibinfo {volume} {54}},\ \bibinfo
  {pages} {715} (\bibinfo {year} {2005})}\BibitemShut {NoStop}%
\bibitem [{\citenamefont {Lee}\ \emph {et~al.}(2006)\citenamefont {Lee},
  \citenamefont {Nagaosa},\ and\ \citenamefont {Wen}}]{Mott}%
  \BibitemOpen
  \bibfield  {author} {\bibinfo {author} {\bibfnamefont {P.~A.}\ \bibnamefont
  {Lee}}, \bibinfo {author} {\bibfnamefont {N.}~\bibnamefont {Nagaosa}}, \ and\
  \bibinfo {author} {\bibfnamefont {X.-G.}\ \bibnamefont {Wen}},\ }\href
  {\doibase 10.1103/RevModPhys.78.17} {\bibfield  {journal} {\bibinfo
  {journal} {Rev. Mod. Phys.}\ }\textbf {\bibinfo {volume} {78}},\ \bibinfo
  {pages} {17} (\bibinfo {year} {2006})}\BibitemShut {NoStop}%
\bibitem [{\citenamefont {Carbotte}(1990)}]{Carbotte1990}%
  \BibitemOpen
  \bibfield  {author} {\bibinfo {author} {\bibfnamefont {J.~P.}\ \bibnamefont
  {Carbotte}},\ }\href {\doibase 10.1103/RevModPhys.62.1027} {\bibfield
  {journal} {\bibinfo  {journal} {Rev. Mod. Phys.}\ }\textbf {\bibinfo {volume}
  {62}},\ \bibinfo {pages} {1027} (\bibinfo {year} {1990})}\BibitemShut
  {NoStop}%
\bibitem [{\citenamefont {Marsiglio}\ and\ \citenamefont
  {Carbotte}(1991{\natexlab{a}})}]{Marsiglio1991}%
  \BibitemOpen
  \bibfield  {author} {\bibinfo {author} {\bibfnamefont {F.}~\bibnamefont
  {Marsiglio}}\ and\ \bibinfo {author} {\bibfnamefont {J.~P.}\ \bibnamefont
  {Carbotte}},\ }\href {\doibase 10.1103/PhysRevB.43.5355} {\bibfield
  {journal} {\bibinfo  {journal} {Phys. Rev. B}\ }\textbf {\bibinfo {volume}
  {43}},\ \bibinfo {pages} {5355} (\bibinfo {year}
  {1991}{\natexlab{a}})}\BibitemShut {NoStop}%
\bibitem [{\citenamefont {Combescot}(1995)}]{Combescot1995}%
  \BibitemOpen
  \bibfield  {author} {\bibinfo {author} {\bibfnamefont {R.}~\bibnamefont
  {Combescot}},\ }\href {\doibase 10.1103/PhysRevB.51.11625} {\bibfield
  {journal} {\bibinfo  {journal} {Phys. Rev. B}\ }\textbf {\bibinfo {volume}
  {51}},\ \bibinfo {pages} {11625} (\bibinfo {year} {1995})}\BibitemShut
  {NoStop}%
\bibitem [{\citenamefont {Marsiglio}\ and\ \citenamefont
  {Carbotte}(2008)}]{Marsiglio2008}%
  \BibitemOpen
  \bibfield  {author} {\bibinfo {author} {\bibfnamefont {F.}~\bibnamefont
  {Marsiglio}}\ and\ \bibinfo {author} {\bibfnamefont {J.~P.}\ \bibnamefont
  {Carbotte}},\ }\enquote {\bibinfo {title} {Electron-phonon
  superconductivity},}\ in\ \href {\doibase 10.1007/978-3-540-73253-2_3} {\emph
  {\bibinfo {booktitle} {Superconductivity: Conventional and Unconventional
  Superconductors}}},\ \bibinfo {editor} {edited by\ \bibinfo {editor}
  {\bibfnamefont {K.~H.}\ \bibnamefont {Bennemann}}\ and\ \bibinfo {editor}
  {\bibfnamefont {J.~B.}\ \bibnamefont {Ketterson}}}\ (\bibinfo  {publisher}
  {Springer Berlin Heidelberg},\ \bibinfo {address} {Berlin, Heidelberg},\
  \bibinfo {year} {2008})\ pp.\ \bibinfo {pages} {73--162}\BibitemShut
  {NoStop}%
\bibitem [{\citenamefont {Allen}\ and\ \citenamefont {Dynes}(1975)}]{ad}%
  \BibitemOpen
  \bibfield  {author} {\bibinfo {author} {\bibfnamefont {P.~B.}\ \bibnamefont
  {Allen}}\ and\ \bibinfo {author} {\bibfnamefont {R.~C.}\ \bibnamefont
  {Dynes}},\ }\href {\doibase 10.1103/PhysRevB.12.905} {\bibfield  {journal}
  {\bibinfo  {journal} {Phys. Rev. B}\ }\textbf {\bibinfo {volume} {12}},\
  \bibinfo {pages} {905} (\bibinfo {year} {1975})}\BibitemShut {NoStop}%
\bibitem [{\citenamefont {Bergmann}\ and\ \citenamefont
  {Rainer}(1973)}]{others1}%
  \BibitemOpen
  \bibfield  {author} {\bibinfo {author} {\bibfnamefont {G.}~\bibnamefont
  {Bergmann}}\ and\ \bibinfo {author} {\bibfnamefont {D.}~\bibnamefont
  {Rainer}},\ }\href {\doibase 10.1007/BF02351862} {\bibfield  {journal}
  {\bibinfo  {journal} {Zeitschrift f{\"u}r Physik}\ }\textbf {\bibinfo
  {volume} {263}},\ \bibinfo {pages} {59} (\bibinfo {year} {1973})}\BibitemShut
  {NoStop}%
\bibitem [{\citenamefont {Marsiglio}\ and\ \citenamefont
  {Carbotte}(1991{\natexlab{b}})}]{others2}%
  \BibitemOpen
  \bibfield  {author} {\bibinfo {author} {\bibfnamefont {F.}~\bibnamefont
  {Marsiglio}}\ and\ \bibinfo {author} {\bibfnamefont {J.~P.}\ \bibnamefont
  {Carbotte}},\ }\href {\doibase 10.1103/PhysRevB.43.5355} {\bibfield
  {journal} {\bibinfo  {journal} {Phys. Rev. B}\ }\textbf {\bibinfo {volume}
  {43}},\ \bibinfo {pages} {5355} (\bibinfo {year}
  {1991}{\natexlab{b}})}\BibitemShut {NoStop}%
\bibitem [{\citenamefont {Karakozov}\ \emph {et~al.}(1991)\citenamefont
  {Karakozov}, \citenamefont {Maksimov},\ and\ \citenamefont
  {Mikhailovsky}}]{others3}%
  \BibitemOpen
  \bibfield  {author} {\bibinfo {author} {\bibfnamefont {A.}~\bibnamefont
  {Karakozov}}, \bibinfo {author} {\bibfnamefont {E.}~\bibnamefont {Maksimov}},
  \ and\ \bibinfo {author} {\bibfnamefont {A.}~\bibnamefont {Mikhailovsky}},\
  }\href {\doibase https://doi.org/10.1016/0038-1098(91)90556-B} {\bibfield
  {journal} {\bibinfo  {journal} {Solid State Communications}\ }\textbf
  {\bibinfo {volume} {79}},\ \bibinfo {pages} {329 } (\bibinfo {year}
  {1991})}\BibitemShut {NoStop}%
\bibitem [{\citenamefont {Bonesteel}\ \emph {et~al.}(1996)\citenamefont
  {Bonesteel}, \citenamefont {McDonald},\ and\ \citenamefont
  {Nayak}}]{Bonesteel1996}%
  \BibitemOpen
  \bibfield  {author} {\bibinfo {author} {\bibfnamefont {N.~E.}\ \bibnamefont
  {Bonesteel}}, \bibinfo {author} {\bibfnamefont {I.~A.}\ \bibnamefont
  {McDonald}}, \ and\ \bibinfo {author} {\bibfnamefont {C.}~\bibnamefont
  {Nayak}},\ }\href {\doibase 10.1103/PhysRevLett.77.3009} {\bibfield
  {journal} {\bibinfo  {journal} {Phys. Rev. Lett.}\ }\textbf {\bibinfo
  {volume} {77}},\ \bibinfo {pages} {3009} (\bibinfo {year}
  {1996})}\BibitemShut {NoStop}%
\bibitem [{\citenamefont {Abanov}\ \emph {et~al.}(2001)\citenamefont {Abanov},
  \citenamefont {Chubukov},\ and\ \citenamefont {Finkel'stein}}]{Abanov2001}%
  \BibitemOpen
  \bibfield  {author} {\bibinfo {author} {\bibfnamefont {A.}~\bibnamefont
  {Abanov}}, \bibinfo {author} {\bibfnamefont {A.~V.}\ \bibnamefont
  {Chubukov}}, \ and\ \bibinfo {author} {\bibfnamefont {A.~M.}\ \bibnamefont
  {Finkel'stein}},\ }\href {http://stacks.iop.org/0295-5075/54/i=4/a=488}
  {\bibfield  {journal} {\bibinfo  {journal} {EPL (Europhysics Letters)}\
  }\textbf {\bibinfo {volume} {54}},\ \bibinfo {pages} {488} (\bibinfo {year}
  {2001})}\BibitemShut {NoStop}%
\bibitem [{\citenamefont {Abanov}\ \emph {et~al.}(2003)\citenamefont {Abanov},
  \citenamefont {Chubukov},\ and\ \citenamefont {Schmalian}}]{Abanov2003}%
  \BibitemOpen
  \bibfield  {author} {\bibinfo {author} {\bibfnamefont {A.}~\bibnamefont
  {Abanov}}, \bibinfo {author} {\bibfnamefont {A.~V.}\ \bibnamefont
  {Chubukov}}, \ and\ \bibinfo {author} {\bibfnamefont {J.}~\bibnamefont
  {Schmalian}},\ }\href {\doibase 10.1080/0001873021000057123} {\bibfield
  {journal} {\bibinfo  {journal} {Advances in Physics}\ }\textbf {\bibinfo
  {volume} {52}},\ \bibinfo {pages} {119} (\bibinfo {year} {2003})}\BibitemShut
  {NoStop}%
\bibitem [{\citenamefont {Son}(1999)}]{son1999}%
  \BibitemOpen
  \bibfield  {author} {\bibinfo {author} {\bibfnamefont {D.~T.}\ \bibnamefont
  {Son}},\ }\href {\doibase 10.1103/PhysRevD.59.094019} {\bibfield  {journal}
  {\bibinfo  {journal} {Phys. Rev. D}\ }\textbf {\bibinfo {volume} {59}},\
  \bibinfo {pages} {094019} (\bibinfo {year} {1999})}\BibitemShut {NoStop}%
\bibitem [{\citenamefont {Chubukov}\ and\ \citenamefont
  {Schmalian}(2005)}]{Chubukov2005}%
  \BibitemOpen
  \bibfield  {author} {\bibinfo {author} {\bibfnamefont {A.~V.}\ \bibnamefont
  {Chubukov}}\ and\ \bibinfo {author} {\bibfnamefont {J.}~\bibnamefont
  {Schmalian}},\ }\href {\doibase 10.1103/PhysRevB.72.174520} {\bibfield
  {journal} {\bibinfo  {journal} {Phys. Rev. B}\ }\textbf {\bibinfo {volume}
  {72}},\ \bibinfo {pages} {174520} (\bibinfo {year} {2005})}\BibitemShut
  {NoStop}%
\bibitem [{\citenamefont {Lee}(2009)}]{sslee}%
  \BibitemOpen
  \bibfield  {author} {\bibinfo {author} {\bibfnamefont {S.-S.}\ \bibnamefont
  {Lee}},\ }\href {\doibase 10.1103/PhysRevB.80.165102} {\bibfield  {journal}
  {\bibinfo  {journal} {Phys. Rev. B}\ }\textbf {\bibinfo {volume} {80}},\
  \bibinfo {pages} {165102} (\bibinfo {year} {2009})}\BibitemShut {NoStop}%
\bibitem [{\citenamefont {Dalidovich}\ and\ \citenamefont
  {Lee}(2013)}]{sslee2}%
  \BibitemOpen
  \bibfield  {author} {\bibinfo {author} {\bibfnamefont {D.}~\bibnamefont
  {Dalidovich}}\ and\ \bibinfo {author} {\bibfnamefont {S.-S.}\ \bibnamefont
  {Lee}},\ }\href {\doibase 10.1103/PhysRevB.88.245106} {\bibfield  {journal}
  {\bibinfo  {journal} {Phys. Rev. B}\ }\textbf {\bibinfo {volume} {88}},\
  \bibinfo {pages} {245106} (\bibinfo {year} {2013})}\BibitemShut {NoStop}%
\bibitem [{\citenamefont {Sachdev}\ \emph {et~al.}(2009)\citenamefont
  {Sachdev}, \citenamefont {Metlitski}, \citenamefont {Qi},\ and\ \citenamefont
  {Xu}}]{subir}%
  \BibitemOpen
  \bibfield  {author} {\bibinfo {author} {\bibfnamefont {S.}~\bibnamefont
  {Sachdev}}, \bibinfo {author} {\bibfnamefont {M.~A.}\ \bibnamefont
  {Metlitski}}, \bibinfo {author} {\bibfnamefont {Y.}~\bibnamefont {Qi}}, \
  and\ \bibinfo {author} {\bibfnamefont {C.}~\bibnamefont {Xu}},\ }\href
  {\doibase 10.1103/PhysRevB.80.155129} {\bibfield  {journal} {\bibinfo
  {journal} {Phys. Rev. B}\ }\textbf {\bibinfo {volume} {80}},\ \bibinfo
  {pages} {155129} (\bibinfo {year} {2009})}\BibitemShut {NoStop}%
\bibitem [{\citenamefont {Moon}\ and\ \citenamefont {Sachdev}(2009)}]{subir2}%
  \BibitemOpen
  \bibfield  {author} {\bibinfo {author} {\bibfnamefont {E.~G.}\ \bibnamefont
  {Moon}}\ and\ \bibinfo {author} {\bibfnamefont {S.}~\bibnamefont {Sachdev}},\
  }\href {\doibase 10.1103/PhysRevB.80.035117} {\bibfield  {journal} {\bibinfo
  {journal} {Phys. Rev. B}\ }\textbf {\bibinfo {volume} {80}},\ \bibinfo
  {pages} {035117} (\bibinfo {year} {2009})}\BibitemShut {NoStop}%
\bibitem [{\citenamefont {Moon}\ and\ \citenamefont {Chubukov}(2010)}]{moon_2}%
  \BibitemOpen
  \bibfield  {author} {\bibinfo {author} {\bibfnamefont {E.-G.}\ \bibnamefont
  {Moon}}\ and\ \bibinfo {author} {\bibfnamefont {A.}~\bibnamefont
  {Chubukov}},\ }\href {\doibase 10.1007/s10909-010-0199-y} {\bibfield
  {journal} {\bibinfo  {journal} {Journal of Low Temperature Physics}\ }\textbf
  {\bibinfo {volume} {161}},\ \bibinfo {pages} {263} (\bibinfo {year}
  {2010})}\BibitemShut {NoStop}%
\bibitem [{\citenamefont {Metlitski}\ and\ \citenamefont
  {Sachdev}(2010{\natexlab{a}})}]{max}%
  \BibitemOpen
  \bibfield  {author} {\bibinfo {author} {\bibfnamefont {M.~A.}\ \bibnamefont
  {Metlitski}}\ and\ \bibinfo {author} {\bibfnamefont {S.}~\bibnamefont
  {Sachdev}},\ }\href {\doibase 10.1103/PhysRevB.82.075127} {\bibfield
  {journal} {\bibinfo  {journal} {Phys. Rev. B}\ }\textbf {\bibinfo {volume}
  {82}},\ \bibinfo {pages} {075127} (\bibinfo {year}
  {2010}{\natexlab{a}})}\BibitemShut {NoStop}%
\bibitem [{\citenamefont {Metlitski}\ and\ \citenamefont
  {Sachdev}(2010{\natexlab{b}})}]{max2}%
  \BibitemOpen
  \bibfield  {author} {\bibinfo {author} {\bibfnamefont {M.~A.}\ \bibnamefont
  {Metlitski}}\ and\ \bibinfo {author} {\bibfnamefont {S.}~\bibnamefont
  {Sachdev}},\ }\href {\doibase 10.1103/PhysRevB.82.075128} {\bibfield
  {journal} {\bibinfo  {journal} {Phys. Rev. B}\ }\textbf {\bibinfo {volume}
  {82}},\ \bibinfo {pages} {075128} (\bibinfo {year}
  {2010}{\natexlab{b}})}\BibitemShut {NoStop}%
\bibitem [{\citenamefont {Mross}\ \emph {et~al.}(2010)\citenamefont {Mross},
  \citenamefont {McGreevy}, \citenamefont {Liu},\ and\ \citenamefont
  {Senthil}}]{senthil}%
  \BibitemOpen
  \bibfield  {author} {\bibinfo {author} {\bibfnamefont {D.~F.}\ \bibnamefont
  {Mross}}, \bibinfo {author} {\bibfnamefont {J.}~\bibnamefont {McGreevy}},
  \bibinfo {author} {\bibfnamefont {H.}~\bibnamefont {Liu}}, \ and\ \bibinfo
  {author} {\bibfnamefont {T.}~\bibnamefont {Senthil}},\ }\href {\doibase
  10.1103/PhysRevB.82.045121} {\bibfield  {journal} {\bibinfo  {journal} {Phys.
  Rev. B}\ }\textbf {\bibinfo {volume} {82}},\ \bibinfo {pages} {045121}
  (\bibinfo {year} {2010})}\BibitemShut {NoStop}%
\bibitem [{\citenamefont {Mahajan}\ \emph {et~al.}(2013)\citenamefont
  {Mahajan}, \citenamefont {Ramirez}, \citenamefont {Kachru},\ and\
  \citenamefont {Raghu}}]{raghu}%
  \BibitemOpen
  \bibfield  {author} {\bibinfo {author} {\bibfnamefont {R.}~\bibnamefont
  {Mahajan}}, \bibinfo {author} {\bibfnamefont {D.~M.}\ \bibnamefont
  {Ramirez}}, \bibinfo {author} {\bibfnamefont {S.}~\bibnamefont {Kachru}}, \
  and\ \bibinfo {author} {\bibfnamefont {S.}~\bibnamefont {Raghu}},\ }\href
  {\doibase 10.1103/PhysRevB.88.115116} {\bibfield  {journal} {\bibinfo
  {journal} {Phys. Rev. B}\ }\textbf {\bibinfo {volume} {88}},\ \bibinfo
  {pages} {115116} (\bibinfo {year} {2013})}\BibitemShut {NoStop}%
\bibitem [{\citenamefont {Fitzpatrick}\ \emph {et~al.}(2013)\citenamefont
  {Fitzpatrick}, \citenamefont {Kachru}, \citenamefont {Kaplan},\ and\
  \citenamefont {Raghu}}]{raghu2}%
  \BibitemOpen
  \bibfield  {author} {\bibinfo {author} {\bibfnamefont {A.~L.}\ \bibnamefont
  {Fitzpatrick}}, \bibinfo {author} {\bibfnamefont {S.}~\bibnamefont {Kachru}},
  \bibinfo {author} {\bibfnamefont {J.}~\bibnamefont {Kaplan}}, \ and\ \bibinfo
  {author} {\bibfnamefont {S.}~\bibnamefont {Raghu}},\ }\href {\doibase
  10.1103/PhysRevB.88.125116} {\bibfield  {journal} {\bibinfo  {journal} {Phys.
  Rev. B}\ }\textbf {\bibinfo {volume} {88}},\ \bibinfo {pages} {125116}
  (\bibinfo {year} {2013})}\BibitemShut {NoStop}%
\bibitem [{\citenamefont {Fitzpatrick}\ \emph {et~al.}(2014)\citenamefont
  {Fitzpatrick}, \citenamefont {Kachru}, \citenamefont {Kaplan},\ and\
  \citenamefont {Raghu}}]{raghu3}%
  \BibitemOpen
  \bibfield  {author} {\bibinfo {author} {\bibfnamefont {A.~L.}\ \bibnamefont
  {Fitzpatrick}}, \bibinfo {author} {\bibfnamefont {S.}~\bibnamefont {Kachru}},
  \bibinfo {author} {\bibfnamefont {J.}~\bibnamefont {Kaplan}}, \ and\ \bibinfo
  {author} {\bibfnamefont {S.}~\bibnamefont {Raghu}},\ }\href {\doibase
  10.1103/PhysRevB.89.165114} {\bibfield  {journal} {\bibinfo  {journal} {Phys.
  Rev. B}\ }\textbf {\bibinfo {volume} {89}},\ \bibinfo {pages} {165114}
  (\bibinfo {year} {2014})}\BibitemShut {NoStop}%
\bibitem [{\citenamefont {Torroba}\ and\ \citenamefont {Wang}(2014)}]{raghu4}%
  \BibitemOpen
  \bibfield  {author} {\bibinfo {author} {\bibfnamefont {G.}~\bibnamefont
  {Torroba}}\ and\ \bibinfo {author} {\bibfnamefont {H.}~\bibnamefont {Wang}},\
  }\href {\doibase 10.1103/PhysRevB.90.165144} {\bibfield  {journal} {\bibinfo
  {journal} {Phys. Rev. B}\ }\textbf {\bibinfo {volume} {90}},\ \bibinfo
  {pages} {165144} (\bibinfo {year} {2014})}\BibitemShut {NoStop}%
\bibitem [{\citenamefont {Fitzpatrick}\ \emph {et~al.}(2015)\citenamefont
  {Fitzpatrick}, \citenamefont {Torroba},\ and\ \citenamefont {Wang}}]{raghu5}%
  \BibitemOpen
  \bibfield  {author} {\bibinfo {author} {\bibfnamefont {A.~L.}\ \bibnamefont
  {Fitzpatrick}}, \bibinfo {author} {\bibfnamefont {G.}~\bibnamefont
  {Torroba}}, \ and\ \bibinfo {author} {\bibfnamefont {H.}~\bibnamefont
  {Wang}},\ }\href {\doibase 10.1103/PhysRevB.91.195135} {\bibfield  {journal}
  {\bibinfo  {journal} {Phys. Rev. B}\ }\textbf {\bibinfo {volume} {91}},\
  \bibinfo {pages} {195135} (\bibinfo {year} {2015})},\ \bibinfo {note} {and
  references therein.}\BibitemShut {Stop}%
\bibitem [{\citenamefont {Fradkin}\ \emph {et~al.}(2010)\citenamefont
  {Fradkin}, \citenamefont {Kivelson}, \citenamefont {Lawler}, \citenamefont
  {Eisenstein},\ and\ \citenamefont {Mackenzie}}]{mack}%
  \BibitemOpen
  \bibfield  {author} {\bibinfo {author} {\bibfnamefont {E.}~\bibnamefont
  {Fradkin}}, \bibinfo {author} {\bibfnamefont {S.~A.}\ \bibnamefont
  {Kivelson}}, \bibinfo {author} {\bibfnamefont {M.~J.}\ \bibnamefont
  {Lawler}}, \bibinfo {author} {\bibfnamefont {J.~P.}\ \bibnamefont
  {Eisenstein}}, \ and\ \bibinfo {author} {\bibfnamefont {A.~P.}\ \bibnamefont
  {Mackenzie}},\ }\href {\doibase 10.1146/annurev-conmatphys-070909-103925}
  {\bibfield  {journal} {\bibinfo  {journal} {Annual Review of Condensed Matter
  Physics}\ }\textbf {\bibinfo {volume} {1}},\ \bibinfo {pages} {153} (\bibinfo
  {year} {2010})}\BibitemShut {NoStop}%
\bibitem [{\citenamefont {Monthoux}\ \emph {et~al.}(2007)\citenamefont
  {Monthoux}, \citenamefont {Pines},\ and\ \citenamefont {Lonzarich}}]{scal}%
  \BibitemOpen
  \bibfield  {author} {\bibinfo {author} {\bibfnamefont {P.}~\bibnamefont
  {Monthoux}}, \bibinfo {author} {\bibfnamefont {D.}~\bibnamefont {Pines}}, \
  and\ \bibinfo {author} {\bibfnamefont {G.~G.}\ \bibnamefont {Lonzarich}},\
  }\href {\doibase 10.1038/nature06480} {\bibfield  {journal} {\bibinfo
  {journal} {Nature}\ }\textbf {\bibinfo {volume} {450}},\ \bibinfo {pages}
  {1177} (\bibinfo {year} {2007})}\BibitemShut {NoStop}%
\bibitem [{\citenamefont {Scalapino}(2012)}]{scal2}%
  \BibitemOpen
  \bibfield  {author} {\bibinfo {author} {\bibfnamefont {D.~J.}\ \bibnamefont
  {Scalapino}},\ }\href {\doibase 10.1103/RevModPhys.84.1383} {\bibfield
  {journal} {\bibinfo  {journal} {Rev. Mod. Phys.}\ }\textbf {\bibinfo {volume}
  {84}},\ \bibinfo {pages} {1383} (\bibinfo {year} {2012})}\BibitemShut
  {NoStop}%
\bibitem [{\citenamefont {Metlitski}\ \emph {et~al.}(2015)\citenamefont
  {Metlitski}, \citenamefont {Mross}, \citenamefont {Sachdev},\ and\
  \citenamefont {Senthil}}]{max_last}%
  \BibitemOpen
  \bibfield  {author} {\bibinfo {author} {\bibfnamefont {M.~A.}\ \bibnamefont
  {Metlitski}}, \bibinfo {author} {\bibfnamefont {D.~F.}\ \bibnamefont
  {Mross}}, \bibinfo {author} {\bibfnamefont {S.}~\bibnamefont {Sachdev}}, \
  and\ \bibinfo {author} {\bibfnamefont {T.}~\bibnamefont {Senthil}},\ }\href
  {\doibase 10.1103/PhysRevB.91.115111} {\bibfield  {journal} {\bibinfo
  {journal} {Phys. Rev. B}\ }\textbf {\bibinfo {volume} {91}},\ \bibinfo
  {pages} {115111} (\bibinfo {year} {2015})}\BibitemShut {NoStop}%
\bibitem [{\citenamefont {Meier}\ \emph {et~al.}(2014)\citenamefont {Meier},
  \citenamefont {P\'epin}, \citenamefont {Einenkel},\ and\ \citenamefont
  {Efetov}}]{efetov}%
  \BibitemOpen
  \bibfield  {author} {\bibinfo {author} {\bibfnamefont {H.}~\bibnamefont
  {Meier}}, \bibinfo {author} {\bibfnamefont {C.}~\bibnamefont {P\'epin}},
  \bibinfo {author} {\bibfnamefont {M.}~\bibnamefont {Einenkel}}, \ and\
  \bibinfo {author} {\bibfnamefont {K.~B.}\ \bibnamefont {Efetov}},\ }\href
  {\doibase 10.1103/PhysRevB.89.195115} {\bibfield  {journal} {\bibinfo
  {journal} {Phys. Rev. B}\ }\textbf {\bibinfo {volume} {89}},\ \bibinfo
  {pages} {195115} (\bibinfo {year} {2014})}\BibitemShut {NoStop}%
\bibitem [{\citenamefont {Efetov}(2015)}]{efetov2}%
  \BibitemOpen
  \bibfield  {author} {\bibinfo {author} {\bibfnamefont {K.~B.}\ \bibnamefont
  {Efetov}},\ }\href {\doibase 10.1103/PhysRevB.91.045110} {\bibfield
  {journal} {\bibinfo  {journal} {Phys. Rev. B}\ }\textbf {\bibinfo {volume}
  {91}},\ \bibinfo {pages} {045110} (\bibinfo {year} {2015})}\BibitemShut
  {NoStop}%
\bibitem [{\citenamefont {Raghu}\ \emph {et~al.}(2015)\citenamefont {Raghu},
  \citenamefont {Torroba},\ and\ \citenamefont {Wang}}]{raghu_15}%
  \BibitemOpen
  \bibfield  {author} {\bibinfo {author} {\bibfnamefont {S.}~\bibnamefont
  {Raghu}}, \bibinfo {author} {\bibfnamefont {G.}~\bibnamefont {Torroba}}, \
  and\ \bibinfo {author} {\bibfnamefont {H.}~\bibnamefont {Wang}},\ }\href
  {\doibase 10.1103/PhysRevB.92.205104} {\bibfield  {journal} {\bibinfo
  {journal} {Phys. Rev. B}\ }\textbf {\bibinfo {volume} {92}},\ \bibinfo
  {pages} {205104} (\bibinfo {year} {2015})}\BibitemShut {NoStop}%
\bibitem [{\citenamefont {Lederer}\ \emph {et~al.}(2015)\citenamefont
  {Lederer}, \citenamefont {Schattner}, \citenamefont {Berg},\ and\
  \citenamefont {Kivelson}}]{steve_sam}%
  \BibitemOpen
  \bibfield  {author} {\bibinfo {author} {\bibfnamefont {S.}~\bibnamefont
  {Lederer}}, \bibinfo {author} {\bibfnamefont {Y.}~\bibnamefont {Schattner}},
  \bibinfo {author} {\bibfnamefont {E.}~\bibnamefont {Berg}}, \ and\ \bibinfo
  {author} {\bibfnamefont {S.~A.}\ \bibnamefont {Kivelson}},\ }\href {\doibase
  10.1103/PhysRevLett.114.097001} {\bibfield  {journal} {\bibinfo  {journal}
  {Phys. Rev. Lett.}\ }\textbf {\bibinfo {volume} {114}},\ \bibinfo {pages}
  {097001} (\bibinfo {year} {2015})}\BibitemShut {NoStop}%
\bibitem [{\citenamefont {Bok}\ \emph {et~al.}(2016)\citenamefont {Bok},
  \citenamefont {Bae}, \citenamefont {Choi}, \citenamefont {Varma},
  \citenamefont {Zhang}, \citenamefont {He}, \citenamefont {Zhang},
  \citenamefont {Yu},\ and\ \citenamefont {Zhou}}]{varma}%
  \BibitemOpen
  \bibfield  {author} {\bibinfo {author} {\bibfnamefont {J.~M.}\ \bibnamefont
  {Bok}}, \bibinfo {author} {\bibfnamefont {J.~J.}\ \bibnamefont {Bae}},
  \bibinfo {author} {\bibfnamefont {H.-Y.}\ \bibnamefont {Choi}}, \bibinfo
  {author} {\bibfnamefont {C.~M.}\ \bibnamefont {Varma}}, \bibinfo {author}
  {\bibfnamefont {W.}~\bibnamefont {Zhang}}, \bibinfo {author} {\bibfnamefont
  {J.}~\bibnamefont {He}}, \bibinfo {author} {\bibfnamefont {Y.}~\bibnamefont
  {Zhang}}, \bibinfo {author} {\bibfnamefont {L.}~\bibnamefont {Yu}}, \ and\
  \bibinfo {author} {\bibfnamefont {X.~J.}\ \bibnamefont {Zhou}},\ }\href
  {\doibase 10.1126/sciadv.1501329} {\bibfield  {journal} {\bibinfo  {journal}
  {Science Advances}\ }\textbf {\bibinfo {volume} {2}} (\bibinfo {year}
  {2016}),\ 10.1126/sciadv.1501329}\BibitemShut {NoStop}%
\bibitem [{\citenamefont {Hartnoll}\ \emph {et~al.}(2011)\citenamefont
  {Hartnoll}, \citenamefont {Hofman}, \citenamefont {Metlitski},\ and\
  \citenamefont {Sachdev}}]{max_2}%
  \BibitemOpen
  \bibfield  {author} {\bibinfo {author} {\bibfnamefont {S.~A.}\ \bibnamefont
  {Hartnoll}}, \bibinfo {author} {\bibfnamefont {D.~M.}\ \bibnamefont
  {Hofman}}, \bibinfo {author} {\bibfnamefont {M.~A.}\ \bibnamefont
  {Metlitski}}, \ and\ \bibinfo {author} {\bibfnamefont {S.}~\bibnamefont
  {Sachdev}},\ }\href {\doibase 10.1103/PhysRevB.84.125115} {\bibfield
  {journal} {\bibinfo  {journal} {Phys. Rev. B}\ }\textbf {\bibinfo {volume}
  {84}},\ \bibinfo {pages} {125115} (\bibinfo {year} {2011})}\BibitemShut
  {NoStop}%
\bibitem [{\citenamefont {Hartnoll}\ \emph {et~al.}(2014)\citenamefont
  {Hartnoll}, \citenamefont {Mahajan}, \citenamefont {Punk},\ and\
  \citenamefont {Sachdev}}]{max_22}%
  \BibitemOpen
  \bibfield  {author} {\bibinfo {author} {\bibfnamefont {S.~A.}\ \bibnamefont
  {Hartnoll}}, \bibinfo {author} {\bibfnamefont {R.}~\bibnamefont {Mahajan}},
  \bibinfo {author} {\bibfnamefont {M.}~\bibnamefont {Punk}}, \ and\ \bibinfo
  {author} {\bibfnamefont {S.}~\bibnamefont {Sachdev}},\ }\href {\doibase
  10.1103/PhysRevB.89.155130} {\bibfield  {journal} {\bibinfo  {journal} {Phys.
  Rev. B}\ }\textbf {\bibinfo {volume} {89}},\ \bibinfo {pages} {155130}
  (\bibinfo {year} {2014})}\BibitemShut {NoStop}%
\bibitem [{\citenamefont {Chubukov}\ \emph {et~al.}(2014)\citenamefont
  {Chubukov}, \citenamefont {Maslov},\ and\ \citenamefont {Yudson}}]{max_23}%
  \BibitemOpen
  \bibfield  {author} {\bibinfo {author} {\bibfnamefont {A.~V.}\ \bibnamefont
  {Chubukov}}, \bibinfo {author} {\bibfnamefont {D.~L.}\ \bibnamefont
  {Maslov}}, \ and\ \bibinfo {author} {\bibfnamefont {V.~I.}\ \bibnamefont
  {Yudson}},\ }\href {\doibase 10.1103/PhysRevB.89.155126} {\bibfield
  {journal} {\bibinfo  {journal} {Phys. Rev. B}\ }\textbf {\bibinfo {volume}
  {89}},\ \bibinfo {pages} {155126} (\bibinfo {year} {2014})}\BibitemShut
  {NoStop}%
\bibitem [{\citenamefont {Maier}\ and\ \citenamefont {Strack}(2016)}]{strack}%
  \BibitemOpen
  \bibfield  {author} {\bibinfo {author} {\bibfnamefont {S.~A.}\ \bibnamefont
  {Maier}}\ and\ \bibinfo {author} {\bibfnamefont {P.}~\bibnamefont {Strack}},\
  }\href {\doibase 10.1103/PhysRevB.93.165114} {\bibfield  {journal} {\bibinfo
  {journal} {Phys. Rev. B}\ }\textbf {\bibinfo {volume} {93}},\ \bibinfo
  {pages} {165114} (\bibinfo {year} {2016})}\BibitemShut {NoStop}%
\bibitem [{\citenamefont {Ridgway}\ and\ \citenamefont
  {Hooley}(2015)}]{strack2}%
  \BibitemOpen
  \bibfield  {author} {\bibinfo {author} {\bibfnamefont {S.~P.}\ \bibnamefont
  {Ridgway}}\ and\ \bibinfo {author} {\bibfnamefont {C.~A.}\ \bibnamefont
  {Hooley}},\ }\href {\doibase 10.1103/PhysRevLett.114.226404} {\bibfield
  {journal} {\bibinfo  {journal} {Phys. Rev. Lett.}\ }\textbf {\bibinfo
  {volume} {114}},\ \bibinfo {pages} {226404} (\bibinfo {year}
  {2015})}\BibitemShut {NoStop}%
\bibitem [{\citenamefont {Haslinger}\ and\ \citenamefont
  {Chubukov}(2003)}]{Haslinger2003}%
  \BibitemOpen
  \bibfield  {author} {\bibinfo {author} {\bibfnamefont {R.}~\bibnamefont
  {Haslinger}}\ and\ \bibinfo {author} {\bibfnamefont {A.~V.}\ \bibnamefont
  {Chubukov}},\ }\href {\doibase 10.1103/PhysRevB.68.214508} {\bibfield
  {journal} {\bibinfo  {journal} {Phys. Rev. B}\ }\textbf {\bibinfo {volume}
  {68}},\ \bibinfo {pages} {214508} (\bibinfo {year} {2003})}\BibitemShut
  {NoStop}%
\bibitem [{\citenamefont {Eliashberg}(1960)}]{Eliashberg}%
  \BibitemOpen
  \bibfield  {author} {\bibinfo {author} {\bibfnamefont {G.~M.}\ \bibnamefont
  {Eliashberg}},\ }\href
  {http://www.jetp.ac.ru/cgi-bin/e/index/e/11/3/p696?a=list} {\bibfield
  {journal} {\bibinfo  {journal} {JETP}\ }\textbf {\bibinfo {volume} {11}},\
  \bibinfo {pages} {696} (\bibinfo {year} {1960})}\BibitemShut {NoStop}%
\bibitem [{xxx()}]{xxxx}%
  \BibitemOpen
  \href@noop {} {\ }\bibinfo {note} {In several $D=2$ systems the corrections
  to Eliashberg formula for the self-energy are singular at the lowest
  frequencies. In this case, Eliashberg theory is valid if $T_c$ is higher than
  a typical frequency at which these singular raghucorrections become
  regular.}\BibitemShut {Stop}%
\bibitem [{com()}]{comm_new}%
  \BibitemOpen
  \href@noop {} {\ }\bibinfo {note} {In the cases, when the interaction is
  peaked for fermions in hot regions on the Fermi surface, the dependence of
  the pairing vertex and the self-energy on the angle $\theta$ along the Fermi
  surface is coupled to frequency dependence, and, strictly speaking, one has
  to solve the set of two 2D coupled integral equations for $\Phi
  (\omega_m,\theta)$ and $\Sigma (\omega_m,\theta)$. This does not lead to
  qualitative changes in the $T_c$ value compared to the corresponding $\gamma$
  models, in which $\Phi$ and $\Sigma$ assumed to be independent on $\theta$ at
  frequencies relevant to $T_c$, but may affect $2\Delta/T_c$
  ratio.}\BibitemShut {Stop}%
\bibitem [{\citenamefont {Altshuler}\ \emph {et~al.}(1995)\citenamefont
  {Altshuler}, \citenamefont {Ioffe},\ and\ \citenamefont
  {Millis}}]{Altshuler1995}%
  \BibitemOpen
  \bibfield  {author} {\bibinfo {author} {\bibfnamefont {B.~L.}\ \bibnamefont
  {Altshuler}}, \bibinfo {author} {\bibfnamefont {L.~B.}\ \bibnamefont
  {Ioffe}}, \ and\ \bibinfo {author} {\bibfnamefont {A.~J.}\ \bibnamefont
  {Millis}},\ }\href {\doibase 10.1103/PhysRevB.52.5563} {\bibfield  {journal}
  {\bibinfo  {journal} {Phys. Rev. B}\ }\textbf {\bibinfo {volume} {52}},\
  \bibinfo {pages} {5563} (\bibinfo {year} {1995})}\BibitemShut {NoStop}%
\bibitem [{\citenamefont {Wang}\ and\ \citenamefont
  {Chubukov}(2013)}]{Wang2013}%
  \BibitemOpen
  \bibfield  {author} {\bibinfo {author} {\bibfnamefont {Y.}~\bibnamefont
  {Wang}}\ and\ \bibinfo {author} {\bibfnamefont {A.}~\bibnamefont
  {Chubukov}},\ }\href {\doibase 10.1103/PhysRevB.88.024516} {\bibfield
  {journal} {\bibinfo  {journal} {Phys. Rev. B}\ }\textbf {\bibinfo {volume}
  {88}},\ \bibinfo {pages} {024516} (\bibinfo {year} {2013})}\BibitemShut
  {NoStop}%
\bibitem [{\citenamefont {Bergeron}\ \emph {et~al.}(2012)\citenamefont
  {Bergeron}, \citenamefont {Chowdhury}, \citenamefont {Punk}, \citenamefont
  {Sachdev},\ and\ \citenamefont {Tremblay}}]{Bergeron2012}%
  \BibitemOpen
  \bibfield  {author} {\bibinfo {author} {\bibfnamefont {D.}~\bibnamefont
  {Bergeron}}, \bibinfo {author} {\bibfnamefont {D.}~\bibnamefont {Chowdhury}},
  \bibinfo {author} {\bibfnamefont {M.}~\bibnamefont {Punk}}, \bibinfo {author}
  {\bibfnamefont {S.}~\bibnamefont {Sachdev}}, \ and\ \bibinfo {author}
  {\bibfnamefont {A.-M.~S.}\ \bibnamefont {Tremblay}},\ }\href {\doibase
  10.1103/PhysRevB.86.155123} {\bibfield  {journal} {\bibinfo  {journal} {Phys.
  Rev. B}\ }\textbf {\bibinfo {volume} {86}},\ \bibinfo {pages} {155123}
  (\bibinfo {year} {2012})}\BibitemShut {NoStop}%
\bibitem [{\citenamefont {Stadler}\ \emph {et~al.}(2015)\citenamefont
  {Stadler}, \citenamefont {Yin}, \citenamefont {von Delft}, \citenamefont
  {Kotliar},\ and\ \citenamefont {Weichselbaum}}]{Stadler2015}%
  \BibitemOpen
  \bibfield  {author} {\bibinfo {author} {\bibfnamefont {K.~M.}\ \bibnamefont
  {Stadler}}, \bibinfo {author} {\bibfnamefont {Z.~P.}\ \bibnamefont {Yin}},
  \bibinfo {author} {\bibfnamefont {J.}~\bibnamefont {von Delft}}, \bibinfo
  {author} {\bibfnamefont {G.}~\bibnamefont {Kotliar}}, \ and\ \bibinfo
  {author} {\bibfnamefont {A.}~\bibnamefont {Weichselbaum}},\ }\href {\doibase
  10.1103/PhysRevLett.115.136401} {\bibfield  {journal} {\bibinfo  {journal}
  {Phys. Rev. Lett.}\ }\textbf {\bibinfo {volume} {115}},\ \bibinfo {pages}
  {136401} (\bibinfo {year} {2015})}\BibitemShut {NoStop}%
\bibitem [{\citenamefont {Lee}\ \emph {et~al.}(2018)\citenamefont {Lee},
  \citenamefont {Chubukov}, \citenamefont {Miao},\ and\ \citenamefont
  {Kotliar}}]{Kotliar}%
  \BibitemOpen
  \bibfield  {author} {\bibinfo {author} {\bibfnamefont {T.-H.}\ \bibnamefont
  {Lee}}, \bibinfo {author} {\bibfnamefont {A.}~\bibnamefont {Chubukov}},
  \bibinfo {author} {\bibfnamefont {H.}~\bibnamefont {Miao}}, \ and\ \bibinfo
  {author} {\bibfnamefont {G.}~\bibnamefont {Kotliar}},\ }\href
  {https://arxiv.org/abs/1805.10280} {\bibfield  {journal} {\bibinfo  {journal}
  {arXiv}\ }\textbf {\bibinfo {volume} {1805}},\ \bibinfo {pages} {10280}
  (\bibinfo {year} {2018})}\BibitemShut {NoStop}%
\bibitem [{\citenamefont {Abanov}\ and\ \citenamefont
  {Chubukov}(1999)}]{Abanov1999}%
  \BibitemOpen
  \bibfield  {author} {\bibinfo {author} {\bibfnamefont {A.}~\bibnamefont
  {Abanov}}\ and\ \bibinfo {author} {\bibfnamefont {A.~V.}\ \bibnamefont
  {Chubukov}},\ }\href {\doibase 10.1103/PhysRevLett.83.1652} {\bibfield
  {journal} {\bibinfo  {journal} {Phys. Rev. Lett.}\ }\textbf {\bibinfo
  {volume} {83}},\ \bibinfo {pages} {1652} (\bibinfo {year}
  {1999})}\BibitemShut {NoStop}%
\bibitem [{\citenamefont {Eschrig}(2006)}]{Eschrig2006}%
  \BibitemOpen
  \bibfield  {author} {\bibinfo {author} {\bibfnamefont {M.}~\bibnamefont
  {Eschrig}},\ }\href {\doibase 10.1080/00018730600645636} {\bibfield
  {journal} {\bibinfo  {journal} {Advances in Physics}\ }\textbf {\bibinfo
  {volume} {55}},\ \bibinfo {pages} {47} (\bibinfo {year} {2006})}\BibitemShut
  {NoStop}%
\bibitem [{\citenamefont {Millis}\ \emph {et~al.}(1988)\citenamefont {Millis},
  \citenamefont {Sachdev},\ and\ \citenamefont {Varma}}]{Millis1988}%
  \BibitemOpen
  \bibfield  {author} {\bibinfo {author} {\bibfnamefont {A.~J.}\ \bibnamefont
  {Millis}}, \bibinfo {author} {\bibfnamefont {S.}~\bibnamefont {Sachdev}}, \
  and\ \bibinfo {author} {\bibfnamefont {C.~M.}\ \bibnamefont {Varma}},\ }\href
  {\doibase 10.1103/PhysRevB.37.4975} {\bibfield  {journal} {\bibinfo
  {journal} {Phys. Rev. B}\ }\textbf {\bibinfo {volume} {37}},\ \bibinfo
  {pages} {4975} (\bibinfo {year} {1988})}\BibitemShut {NoStop}%
\bibitem [{\citenamefont {Wang}\ \emph {et~al.}(2016)\citenamefont {Wang},
  \citenamefont {Abanov}, \citenamefont {Altshuler}, \citenamefont
  {Yuzbashyan},\ and\ \citenamefont {Chubukov}}]{Wang2016}%
  \BibitemOpen
  \bibfield  {author} {\bibinfo {author} {\bibfnamefont {Y.}~\bibnamefont
  {Wang}}, \bibinfo {author} {\bibfnamefont {A.}~\bibnamefont {Abanov}},
  \bibinfo {author} {\bibfnamefont {B.~L.}\ \bibnamefont {Altshuler}}, \bibinfo
  {author} {\bibfnamefont {E.~A.}\ \bibnamefont {Yuzbashyan}}, \ and\ \bibinfo
  {author} {\bibfnamefont {A.~V.}\ \bibnamefont {Chubukov}},\ }\href {\doibase
  10.1103/PhysRevLett.117.157001} {\bibfield  {journal} {\bibinfo  {journal}
  {Phys. Rev. Lett.}\ }\textbf {\bibinfo {volume} {117}},\ \bibinfo {pages}
  {157001} (\bibinfo {year} {2016})}\BibitemShut {NoStop}%
\bibitem [{\citenamefont {Abanov}\ \emph {et~al.}(2008)\citenamefont {Abanov},
  \citenamefont {Chubukov},\ and\ \citenamefont {Norman}}]{Abanov2008}%
  \BibitemOpen
  \bibfield  {author} {\bibinfo {author} {\bibfnamefont {A.}~\bibnamefont
  {Abanov}}, \bibinfo {author} {\bibfnamefont {A.~V.}\ \bibnamefont
  {Chubukov}}, \ and\ \bibinfo {author} {\bibfnamefont {M.~R.}\ \bibnamefont
  {Norman}},\ }\href {\doibase 10.1103/PhysRevB.78.220507} {\bibfield
  {journal} {\bibinfo  {journal} {Phys. Rev. B}\ }\textbf {\bibinfo {volume}
  {78}},\ \bibinfo {pages} {220507} (\bibinfo {year} {2008})}\BibitemShut
  {NoStop}%
\bibitem [{\citenamefont {Anderson}(1959)}]{ANDERSON195926}%
  \BibitemOpen
  \bibfield  {author} {\bibinfo {author} {\bibfnamefont {P.}~\bibnamefont
  {Anderson}},\ }\href {\doibase https://doi.org/10.1016/0022-3697(59)90036-8}
  {\bibfield  {journal} {\bibinfo  {journal} {Journal of Physics and Chemistry
  of Solids}\ }\textbf {\bibinfo {volume} {11}},\ \bibinfo {pages} {26 }
  (\bibinfo {year} {1959})}\BibitemShut {NoStop}%
\bibitem [{\citenamefont {Wang}\ \emph {et~al.}(2001)\citenamefont {Wang},
  \citenamefont {Mao},\ and\ \citenamefont {Bedell}}]{triplet}%
  \BibitemOpen
  \bibfield  {author} {\bibinfo {author} {\bibfnamefont {Z.}~\bibnamefont
  {Wang}}, \bibinfo {author} {\bibfnamefont {W.}~\bibnamefont {Mao}}, \ and\
  \bibinfo {author} {\bibfnamefont {K.}~\bibnamefont {Bedell}},\ }\href
  {\doibase 10.1103/PhysRevLett.87.257001} {\bibfield  {journal} {\bibinfo
  {journal} {Phys. Rev. Lett.}\ }\textbf {\bibinfo {volume} {87}},\ \bibinfo
  {pages} {257001} (\bibinfo {year} {2001})}\BibitemShut {NoStop}%
\bibitem [{\citenamefont {Roussev}\ and\ \citenamefont
  {Millis}(2001)}]{triplet2}%
  \BibitemOpen
  \bibfield  {author} {\bibinfo {author} {\bibfnamefont {R.}~\bibnamefont
  {Roussev}}\ and\ \bibinfo {author} {\bibfnamefont {A.~J.}\ \bibnamefont
  {Millis}},\ }\href {\doibase 10.1103/PhysRevB.63.140504} {\bibfield
  {journal} {\bibinfo  {journal} {Phys. Rev. B}\ }\textbf {\bibinfo {volume}
  {63}},\ \bibinfo {pages} {140504} (\bibinfo {year} {2001})}\BibitemShut
  {NoStop}%
\bibitem [{\citenamefont {Chubukov}\ \emph {et~al.}(2003)\citenamefont
  {Chubukov}, \citenamefont {Finkel'stein}, \citenamefont {Haslinger},\ and\
  \citenamefont {Morr}}]{triplet3}%
  \BibitemOpen
  \bibfield  {author} {\bibinfo {author} {\bibfnamefont {A.~V.}\ \bibnamefont
  {Chubukov}}, \bibinfo {author} {\bibfnamefont {A.~M.}\ \bibnamefont
  {Finkel'stein}}, \bibinfo {author} {\bibfnamefont {R.}~\bibnamefont
  {Haslinger}}, \ and\ \bibinfo {author} {\bibfnamefont {D.~K.}\ \bibnamefont
  {Morr}},\ }\href {\doibase 10.1103/PhysRevLett.90.077002} {\bibfield
  {journal} {\bibinfo  {journal} {Phys. Rev. Lett.}\ }\textbf {\bibinfo
  {volume} {90}},\ \bibinfo {pages} {077002} (\bibinfo {year}
  {2003})}\BibitemShut {NoStop}%
\bibitem [{\citenamefont {Dolgov}\ \emph {et~al.}(2005)\citenamefont {Dolgov},
  \citenamefont {Mazin}, \citenamefont {Golubov}, \citenamefont {Savrasov},\
  and\ \citenamefont {Maksimov}}]{Dolgov2005}%
  \BibitemOpen
  \bibfield  {author} {\bibinfo {author} {\bibfnamefont {O.~V.}\ \bibnamefont
  {Dolgov}}, \bibinfo {author} {\bibfnamefont {I.~I.}\ \bibnamefont {Mazin}},
  \bibinfo {author} {\bibfnamefont {A.~A.}\ \bibnamefont {Golubov}}, \bibinfo
  {author} {\bibfnamefont {S.~Y.}\ \bibnamefont {Savrasov}}, \ and\ \bibinfo
  {author} {\bibfnamefont {E.~G.}\ \bibnamefont {Maksimov}},\ }\href {\doibase
  10.1103/PhysRevLett.95.257003} {\bibfield  {journal} {\bibinfo  {journal}
  {Phys. Rev. Lett.}\ }\textbf {\bibinfo {volume} {95}},\ \bibinfo {pages}
  {257003} (\bibinfo {year} {2005})}\BibitemShut {NoStop}%
\bibitem [{\citenamefont {Marsiglio}(2018)}]{Marsiglio2018}%
  \BibitemOpen
  \bibfield  {author} {\bibinfo {author} {\bibfnamefont {F.}~\bibnamefont
  {Marsiglio}},\ }\href {\doibase 10.1103/PhysRevB.98.024523} {\bibfield
  {journal} {\bibinfo  {journal} {Phys. Rev. B}\ }\textbf {\bibinfo {volume}
  {98}},\ \bibinfo {pages} {024523} (\bibinfo {year} {2018})}\BibitemShut
  {NoStop}%
\bibitem [{\citenamefont {Chubukov}\ and\ \citenamefont
  {Maslov}(2012)}]{Chubukov2012}%
  \BibitemOpen
  \bibfield  {author} {\bibinfo {author} {\bibfnamefont {A.~V.}\ \bibnamefont
  {Chubukov}}\ and\ \bibinfo {author} {\bibfnamefont {D.~L.}\ \bibnamefont
  {Maslov}},\ }\href {\doibase 10.1103/PhysRevB.86.155136} {\bibfield
  {journal} {\bibinfo  {journal} {Phys. Rev. B}\ }\textbf {\bibinfo {volume}
  {86}},\ \bibinfo {pages} {155136} (\bibinfo {year} {2012})}\BibitemShut
  {NoStop}%
\end{thebibliography}%

\end{document}